\documentclass[a4paper,fleqn,usenatbib,useAMS]{mnras}

\usepackage{color}

\usepackage{graphicx}	
\usepackage{amsmath}	
\usepackage{amssymb}	
\usepackage{multicol}        
\usepackage{bm}		
\usepackage{pdflscape}	

\usepackage{xcolor}

\usepackage[T1]{fontenc}
\usepackage{ae,aecompl}

\title[First-order Fermi acceleration properties in  reconnection]{Properties of the First-order Fermi acceleration in fast magnetic reconnection driven by turbulence {in collisional MHD flows}}

\author[del Valle, de Gouveia Dal Pino \& Kowal]{
Maria V. del Valle,$^{1}$\thanks{E-mail: maria@iar-conicet.gov.ar (MVdV)}
E. M. de Gouveia Dal Pino,$^{2}$\thanks{E-mail: dalpino@iag.usp.br (EMGDP)}
and G. Kowal$^{3,4}$
\\
$^{1}$Instituto Argentino de Radioastronom\'{\i}a (IAR), Camino General Belgrano Km 40, Argentina\\
$^{2}$Instituto de Astronomia, Geof\'\i sica e Ci\^encias Atmosf\'ericas,
             Universidade de S\~ao Paulo, Rua do Mat\~ao 1226, 05508-900, S\~ao Paulo, Brazil\\
$^{3}$N\'{u}cleo de Astrof\'{\i}sica Te\'orica, Universidade Cruzeiro do Sul, Rua Galv\~{a}o Bueno 868, 01506-000, S\~ao Paulo, Brazil\\
$^{4}$Escola de Artes, Ciencias e Humanidades, Universidade de S\~ao Paulo, Rua Arlindo Bettio 1000, 03828-000, S\~ao Paulo, Brazil
}

\pubyear{2016}

\begin{document}
\label{firstpage}
\pagerange{\pageref{firstpage}--\pageref{lastpage}}
\maketitle

\begin{abstract}

Fast magnetic reconnection may occur in different astrophysical sources, producing flare-like emission and particle acceleration. Currently, this process  is being studied as an efficient mechanism to  accelerate particles via a first-order Fermi process.
In this work we  analyse  the acceleration rate and the energy distribution of test particles injected in three-dimensional magnetohydrodynamical (MHD) domains with {large-scale} current sheets where reconnection is made fast by the presence of turbulence.
{We study the dependence of the particle acceleration time with the relevant  parameters of the {embedded turbulence}, i.e., the Alfv\'en  {speed} $V_{\rm A}$, the injection power  $P_{\rm inj}$ and scale $k_{\rm inj}$ ($k_{\rm inj} = 1/l_{\rm inj}$). We find that the acceleration time follows a power-law dependence with the particle {kinetic} energy: $t_{acc} \propto E^{\alpha}$, with $0.2 < \alpha < 0.6$} for a vast range of values of  $c / V_{\rm A} \sim 20 - 1000$.
The acceleration time decreases with the {Alfv\'en speed} {(and therefore with the reconnection velocity)} as expected,   having an approximate dependence  $t_{acc}  \propto (V_{\rm A} / c)^{-\kappa}$, {with $\kappa \sim 2.1- 2.4$ {for particles reaching  kinetic} energies between {$1 - 100 \, m_p c^2$}, respectively. Furthermore, we find that the acceleration time is only weakly dependent on the $P_{\rm inj}$ and $l_{\rm inj}$  parameters} of the turbulence.
 The particle spectrum develops a high-energy tail {which can be fitted by a hard power-law already in the early times of the acceleration, in consistency  with the results of kinetic studies of particle acceleration by magnetic reconnection in collisionless plasmas}.

\end{abstract}




\begingroup
\let\clearpage\relax
\endgroup
\newpage

\graphicspath{{./figs/}}

\section{Introduction}

Magnetic reconnection occurs when two magnetic fluxes of opposite polarity
encounter each other. Under finite magnetic resistivity conditions  a current
sheet is formed at the discontinuity surface, where the field lines annihilate.
Direct evidence of magnetic reconnection in astrophysical and space environments
like the solar corona and the Earth magnetotail indicate that in some
circumstances reconnection can be very fast, with rates which are a substantial
fraction  of the Alfv\'en speed $V_{\rm A}$.

Fast reconnection breaks the magnetic field topology releasing  magnetic energy
explosively thus  explaining the bursty emission, for instance, in  solar
flares. Relativistic particles are always observed in connection with these
flares suggesting that magnetic reconnection can lead to direct particle
acceleration \citep[see e.g., the reviews ][and references therein]{degouveia14, degouveia15, uzdensky2011}.

In analogy to diffusive shock acceleration (DSA), in which particles confined
between the upstream  and downstream  flows undergo a first-order Fermi
acceleration, \citet{degouveia05} (hereafter GL05) proposed  a similar
process  occurring within the current sheet where trapped particles bounce
back and forth between the converging magnetic fluxes of opposite polarity  in the
{large-scale} reconnection region. The particles  gyrorotate around a reconnected
magnetic field \citep[see Figure 2b in][]{kowal11}, gaining energy due to
collisions with magnetic irregularities at a rate $\Delta E/E \propto V_{\rm
rec}/c$ (where $V_{\rm rec}$ is the reconnection speed) {implying} a
first-order Fermi process with an exponential energy growth after several round
trips (GL05, \citealt{degouveia15}).
{A similar process was also invoked by \citet{drake06} who investigated particles accelerated inside  two-dimensional contracting magnetic islands (or loops). In \citet{kowal11} it has been demonstrated the equivalence between the two mechanisms for driving  first-order Fermi acceleration. This process has been extensively tested numerically mainly through two-dimensional (2D) particle-in-cell (PIC) simulations of collisionless electron-ion or electron-positron  plasmas \citep[e.g.,][]{drake06,zenitani01,zenitani07, zenitani08,lyubarsky08,drake10,clausen-brown2012, cerutti14, li15}, and more recently also through three-dimensional (3D) PIC simulations \citep{sironi14,guo15, guo16}. However, these simulations can probe acceleration only at the kinetic scales of the plasma, of a few hundreds of the inertial length   ($\sim 100 c/\omega_p$, where $\omega_p$ is the plasma frequency). To assess the first-order Fermi process in the large scales of the collisional MHD flows commonly observed in astrophysical systems, \citet{kowal11, kowal12a} have also successfully tested it in 2D and 3D MHD simulations injecting test particles in the reconnection domain.
}

Currently, fast magnetic reconnection is regarded as a potentially important
mechanism to accelerate particles not only in the solar system context
\citep[e.g.,][]{drake06, drake09, gordovskyy10, gordovskyy11, zharkova11,
lazarian09, drake10, lazarian10,li15}, but also beyond it, in galactic and
extragalactic environments such as jet-accretion disk systems
\citep[e.g.,][]{degouveia05,degouveia10a,degouveia10b, giannios10,delvalle11,
kadowaki14,khiali15a,khiali15b}, pulsar {winds} and GRBs
\citep[e.g.,][]{lazarian03,zenitani07, zhang11,uzdensky2011,clausen-brown2012,cerutti14,
sironi14,guo14,guo15,singh16}. It has been also related  to the production of {ultra-high-energy}
cosmic rays \citep[e.g.,][]{degouveia00, degouveia01, kotera2011}. Besides, the
accelerated  particles may produce detectable non-thermal emission in a wide
range of energies, specially at gamma rays
\citep[e.g.,][]{delvalle11,vieyro12,cerutti14,khiali15a,khiali15b,kadowaki14,
singh14} or neutrinos \citep[e.g.,][]{khiali16}, {therefore, studies of} the
acceleration rate and the particle power-law index are fundamental for
understanding and modelling this emission.

As remarked above, in order to obtain an efficient acceleration process, reconnection has
to be fast. {In collisioneless plasmas, this is usually ensured by kinetic instabilities
or by the Hall effect (in the case of an electron-ion plasma), both  relevant only at plasma kinetic scales. In {large-scale} collisional MHD systems, fast reconnection can be driven  either by anomalous resistivity \citep[][]{parker79,biskamp_etal97,zenitani09} or by turbulence} \citep[][]{lazarian99, kowal09, kowal12b,lazarian12}\footnote{Alternative descriptions of fast reconnection in a collisional MHD scenario have been proposed also by \citet{loureiro07,shibata01}.}.

In a weak turbulent medium, the wandering of the magnetic field lines  allows
for many simultaneous events of reconnection {to happen at the same time}. {Moreover, the reconnected flux is more
efficiently removed due to turbulence which broadens the outflow channel
\citep[see Figure 1 in][for example]{kowal09}. These two factors make such
reconnection fast.} According to \citet{lazarian99}, $V_{\rm rec} \sim V_{\rm A}
(l/L)^{1/2} (v_l/V_{\rm A})^2$, where $v_l$ and $l$ are the injection
velocity {and scale} of the turbulence, respectively. It is easy to see that for {the upper limit, i.e.} $l
\sim L$ and $v_l \sim V_{\rm A}$, {the maximum reconnection rate is} $V_{\rm rec}
\sim V_{\rm A}$. Both features, the simultaneous reconnection events and the broadened
reconnection layer,  are very important for accelerating particles, as
demonstrated in \citet{kowal11, kowal12a}.

In this work we extend  the earlier numerical studies of \citet{kowal11,
kowal12a}  of the acceleration of test particles in collisional,
non-relativistic\footnote{Recent studies \citep{takamoto15} indicate that in
relativistic domains turbulent driven magnetic reconnection behaves similarly to
the non-relativistic case  \citep[see][for reviews]{lazarian15, degouveia14}.}
three-dimensional MHD domains of reconnection having {large-scale}
current sheets {with embedded turbulence, in order to assess  the
dependence of the particles  acceleration time and power spectrum with the
parameters involved in the process, namely, the  reconnection speed which in
turn is directly correlated with the Alfv\'en velocity ($V_{\rm A}$),  and  the
turbulence injection power ($P_{\rm inj}$) and  scale ($k_{\rm
inj} = 1/l_{\rm inj}$) {using the same methodology as described in
\citet{kowal12a}}.}
\footnote{We note that in an earlier pioneering work,  \citet{kobak2000} also studied the role of MHD turbulence in the particle acceleration process in a volume with a reconnecting magnetic field. However, they  did not consider a real turbulent cascade developed self-consistently to affect the reconnection at the current sheet,  as performed in \citet{kowal11,kowal12a} and  in the present work. They instead, employed a Monte Carlo method and mimicked the effects of the turbulence with  small-amplitude pitch angle scatterings. Their approach did not allow them to detect any Fermi process. Besides, the limitations of their method did not allow them to explore the dependence of the acceleration rate with the parameters of the turbulence, or the Alfv\'en (and reconnection) velocity, as we do in the present work.}

In the next section we  summarize  the main aspects of  the theory of
first-order Fermi acceleration within current sheets with fast  reconnection
driven by turbulence. In Sec.~\ref{sec:methods}, we describe the numerical
methodology used in this work to perform the calculations. In Sec.~\ref{sec:
time} we show the computed acceleration time for different models. In Sec.
~\ref{sec:acc_dist} we analyse the distribution of the accelerated particles. In
 Sec.~\ref{sec:concl} we discuss the results and draw  our conclusions.

\section{Acceleration model}
\label{sec:theory}

As remarked, in this work we explore the first-order Fermi acceleration process
within {large-scale} current sheets in collisional MHD domains with fast
reconnection  driven by embedded turbulence.  The overall aspects of this
mechanism have been thoroughly discussed in several papers and recent reviews
\citep[GL05, ][]{kowal11,kowal12a,degouveia15} and here we just summarize the main
assumptions.

The mechanism of first-order Fermi acceleration operating within {large-scale}
current sheets first proposed by  GL05, with trapped particles  moving back and
forth between the two converging reconnecting magnetic flux tubes, undergoing
collisions with magnetic fluctuations and suffering  a net energy gain
\begin{equation}
 \langle \Delta E / E \rangle \sim  V_{\rm rec}/ c
 \label{eq:energy_gain}
\end{equation}
after each round trip, is completely
general and works either in collisional or collisionless fluids in two and
three-dimension domains \citep{kowal11}. Furthermore, it is  equivalent to the
first-order particle acceleration process within two-dimensional converging
magnetic islands proposed by \citep{drake06}, as demonstrated in \citet{kowal11}.

The expression (\ref{eq:energy_gain}) {indicates that  after several round trips the particle energy must increase exponentially.} {In addition,} it clearly shows that  reconnection has to be fast ($V_{\rm
rec} \sim V_{\rm A}$)  in order to make the overall  process efficient. For
instance, in the surroundings of relativistic sources  $V_{\rm rec} \simeq  V_{\rm A}
\simeq c$
\citep[GL05,][]{degouveia10a, giannios10,
lazarian05}\footnote{\citet{giannios10}, in particular, repeated GL05
calculations for a relativistic MHD flow and  obtained a net particle energy
increase $ \langle \Delta E / E \rangle \sim  4 \beta_{\rm rec} / 3 + \beta_{\rm
rec}^2/2$, where ${\beta}_{\rm rec} = V_{\rm rec} /c $, which in the limit
$\beta_{\rm rec} << 1$ recovers the expression obtained by GL05.}.

Making a simple approximation that  particles would escape from the acceleration zone with a similar
rate as in shock acceleration,  GL05 predicted an analytical  power-law energy
distribution for the accelerated particles $N(E) \sim E^{-5/2}$ which is actually
compatible, e.g., with observed synchrotron power-law spectra slope
in microquasars. Relaxing  the escape assumption, \citet{drury12} found that the
particles may leave the acceleration zone with a similar  rate as in shock
acceleration, which results {in} $N(E) \sim E^{-1}$   if the compression ratio between
the outflow and inflow densities in the reconnection site is large.
It is worth noting that {the Lazarian \& Vishniac (1999)
fast reconnection  is built upon an  incompressible turbulent theory and the numerical simulations employed here which confirm this (Kowal et al. 2009, 2011, 2012a) were done in the nearly incompressible case, where the inflow/outflow density ratio is close to unity. Indeed, the variations of density in our computational domain  are smaller than 5\% and  are very sensitive to the sonic Mach number. Since our sound speed is 4.0, we have  very small sonic Mach number, of the order of 0.25 or smaller. 
Therefore, Drury's (2012) result does not apply in our context and some anisotropy between the parallel and perpendicular components
of the  particle velocities is necessary in order to ensure efficient particle
acceleration. In \citet{kowal11, kowal12a}, it has been found that the anisotropy
arises  from particles preferentially being accelerated either in the parallel
or the perpendicular direction in the beginning of the process 
\citep[see also discussion in][]{degouveia14, degouveia15}. As in \citet{kowal12a}, in Section 3  
 we depict
results of numerical simulations of particle acceleration by magnetic
reconnection in large current sheets in the presence of turbulence where the
evolution of both particle velocity components is tracked separately in order to
further  address this point.

\subsection{Fast reconnection driven by turbulence}

In this work we focus on fast reconnection driven by turbulence as described  by
\citet{lazarian99} \citep[see also][]{eyink2011} and tested
numerically through three-dimensional simulations by \citet{kowal09, kowal12b}.
As mentioned in Introduction, in this  model, the magnetic field wandering is the
key process that induces fast,  independent of  Ohmic resistivity, magnetic
reconnection. The presence of turbulence enables the wandering of the magnetic
lines and therefore, reconnection {events} occurring simultaneously making
reconnection very fast.  For extremely weak turbulence the reconnection rate
reduces to the well known slow Sweet-Parker rate. For  further reading on this
fast reconnection model we refer to the recent reviews by \citet{lazarian12, lazarian14, lazarian15}.

{The predicted} dependence of magnetic reconnection on the properties
of turbulence, i.e. on the injection power $P_{\rm inj}$ and injection scale of
turbulence $l_{\rm inj}$, is given by \citep{lazarian99, kowal09}:
\begin{equation}
\frac{V_{\rm rec}}{{V_{\rm A}}} \propto P_{\rm inj}^{1/2}\,{l_{\rm inj}}.
\label{vrec}
\end{equation}
but the   numerical tests by \citet{kowal09, kowal12b} indicate a weaker dependence on $l_{\rm inj}$, i.e. ${V_{\rm rec}}/{{V_{\rm A}}} \propto P_{\rm inj}^{1/2}\,{l_{\rm inj}}^{3/4}$. In this work we consider {the estimated} relation between  $V_{\rm rec}$ and the turbulence to explore the dependence of the first-order Fermi  particle acceleration  on the turbulent fast reconnection parameters.

\subsection{Particle acceleration within current sheets with fast reconnection driven by turbulence}

In a  Sweet-Parker reconnection configuration with reconnection made
artificially fast by enhanced numerical resistivity, particles are accelerated
through a first-order Fermi process, as predicted in GL05 and demonstrated
numerically in \citet{kowal11} and \citet{kowal12a}.  When turbulence is included within the
current sheet  \citep[see][]{kowal12a} the acceleration {process is improved.} As mentioned
earlier, this is because  the presence of turbulence allows the formation of a
thick volume filled with multiple simultaneously reconnecting magnetic lines,
making the process intrinsically three-dimensional. Charged particles
trapped within this volume suffer several head-on scatterings with the
contracting magnetic fluctuations, which significantly improves the acceleration.

The local effective accelerating electric field is
$(\mathbf{u}-\mathbf{v})\times \mathbf{B}$, where $\mathbf{u}$ is the particle
velocity, $\mathbf{v}$ is the local velocity of {plasma fluctuations}, and
$\mathbf{B}$ is the local magnetic field (see below).

In the next section we briefly describe the numerical method employed to construct the scenario above and inject test particles, as in Kowal et al. (2012a).

\section{Numerical Methodology}
\label{sec:methods}

Following \citet{kowal12a}, we inject test particles (5,000 - 10,000 protons) into a
frozen-in-time  3D  MHD  domain with a {large-scale}
current sheet containing embedded weakly stochastic
turbulence. {This domain is built by integrating numerically the collisional MHD equations, which are appropriate for the description of most macroscopic astrophysical flows, until the turbulence in the current sheet reaches a steady-state.  Considering that the macroscopic MHD dynamical times are much longer than those of the test particles, using  a single snapshot to accelerate the particles is a reasonable assumption (see Sec.~\ref{time-evol}).}

For each {test} particle  {we numerically solve} the
relativistic equation of motion:
\begin{equation}
\frac{d}{dt} (\gamma \, m \, {\bf u}) = q({\bf E} + {\bf u}\times{\bf B}),
\end{equation}
where $m$, $q$, $\gamma$ and ${\bf u}$ are the particle mass, charge, Lorentz factor, and {particle} velocity, respectively. ${\bf E}$ and  ${\bf B}$ are the electric and magnetic fields, respectively. The electric field is obtained from Ohm's law:
\begin{equation}
{\bf E} = -{\bf v}\times {\bf B} + \eta{\bf J},
\end{equation}
where ${\bf v}$ is the flow velocity, ${\bf J}$ is the current density and $\eta$ is the Ohmic resistivity coefficient. The contribution to the electric field due to the resistivity is neglected, since its effects on the particle acceleration {for realistic resistivities are smaller}. Therefore, the equation of motion is
\begin{equation}
\label{eq:mov}
\frac{d}{dt} (\gamma \, m \, {\bf u}) = q[({\bf u}-{\bf v})\times{\bf B}].
\end{equation}
\footnote{We note that the  ''collisional'' term in our study refers only to the macroscopic MHD  flow where magnetic reconnection takes place. The test particles are injected in this flow  and then are accelerated due to interactions with local fluctuations through the effective electric field as described by Eq.~(\ref{eq:mov}).}
This equation is integrated {in time for the particles} using a 4$^{\rm th}$ order Runge-Kutta method, and the fields are interpolated up to second order \citep[see][]{kowal12a}.

Particles {are initiated with} random positions and {velocity} directions and a thermal velocity distribution. The {thermal distribution speed} is fixed to $v_{th}=0.04 \, c$ for all cases. No radiative loss is included, {so} particles lose or gain energy only by the interactions with the plasma {fluctuations}.

\subsection{Initial Conditions}\label{data}

Distinctly from \citet{kowal12a} who considered only one MHD  model of
magnetic reconnection, here we explore different  conditions for the
embedded turbulence in the current sheet. The models studied  are described in
Table~\ref{tabla}.

All models begin with a Harris current sheet $B_{x}(x,y,z)$ $=$
$B_{0x}\tanh(y/\theta)$, initialized using the magnetic vector potential
$A_{z}(x,y,z) = \ln |\cosh(y/\theta)|$, with a uniform guide field $B_{0z}$. The
initial density profile is set from the condition of uniform total pressure, and
the initial velocity is {set} to zero. The reconnection is
initiated by a small perturbation to the magnetic vector potential $\delta
A_{x}(x,y,z)$ = $\delta B_{0x} \cos(2 \pi x)\exp[-(y/d)^2]$, with $\delta B_{0z}
= 0.05$ and $d = 0.1$ (see also \citet{kowal09} for more details).


The velocity and magnetic field are expressed in Alfv\'en speed units, defined
by the antiparallel component of the magnetic field, and the initial
unperturbed density $\rho_{0} = 1$.  The  initial strength of the antiparallel
component of the magnetic field is set to $B_{0x} = 1.0$ in all models, and the
guide field to $B_{0z} = 0.1$.  The distance unit is defined by the length of
the computational  box in the $x$ direction. The box has size $L_{x} = L_{z} =
1$ and $L_{y} = 2$. The time is measured in units of $L_{x}/V_{\rm A}$.  The
computational box has a grid of $256\times512\times256$ for all models. For more
details see  \citet{kowal09} and \citet{kowal12b}.

Turbulence is driven using a method  described in \citet{alvelius99}. The
driving  is made by injecting discrete velocity fluctuations in the Fourier
space concentrated around the wave vector $k_{\rm inj}$, corresponding to the
injection scale $l_{\rm inj}$. The amplitude of driving is solely determined by its power
$P_{\rm inj}$, the number of driven Fourier components, and the time step of
driving. In all models the turbulence is sub-Alfv\'enic. See
\citet{kowal09} for further details. {When} the turbulence {is developed} and reach a steady-sate, we injected
 test particles in this domain  and followed their
trajectories according to Eq.~(\ref{eq:mov}). The turbulence injection power and
length scale for the studied models are given in Table~\ref{tabla}.

{A map illustrating  the trajectory of an injected test particle in the  3D magnetic reconnection domain  is depicted in Figure \ref{trajectory}.} 


\begin{figure}
\begin{centering}
\includegraphics[width=1.\linewidth,angle=0 ]{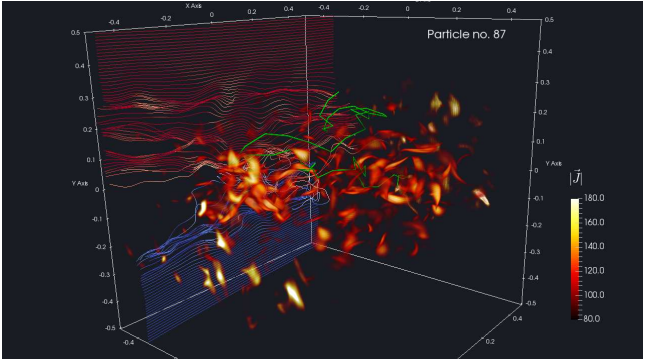}
\caption{This map illustrates the trajectory (in green) of an injected test particle bouncing up and down in the  3D magnetic reconnection domain simulated with  turbulence embedded in it  to drive fast reconnection. The volume depicts the current sheet region  with several very high current density patches where the test particle is scattered by magnetic fluctuations while accelerated. The blue and red lines represent the two magnetic line fluxes of opposite polarity.
}
\label{trajectory}
\end{centering}
\end{figure}

\begin{table}
\caption{{Turbulent MHD reconnection} model parameters, all values are {expressed} in code units.}
\begin{center}
\begin{tabular}{llllll}
\hline
Model & I & II & III & IV & V\\
\hline
$P_{\rm inj}$ & $1$ & $0.5$ &  $0.1$ & $1$  & $1$\\
$k_{\rm inj}$ & $8$ & $8$ &  $8$ & $12$ & $16$\\
\hline
\end{tabular}
\end{center}
\label{tabla}
\end{table}

\begin{figure}
\begin{centering}
\includegraphics[width=0.8\linewidth,angle=0 ]{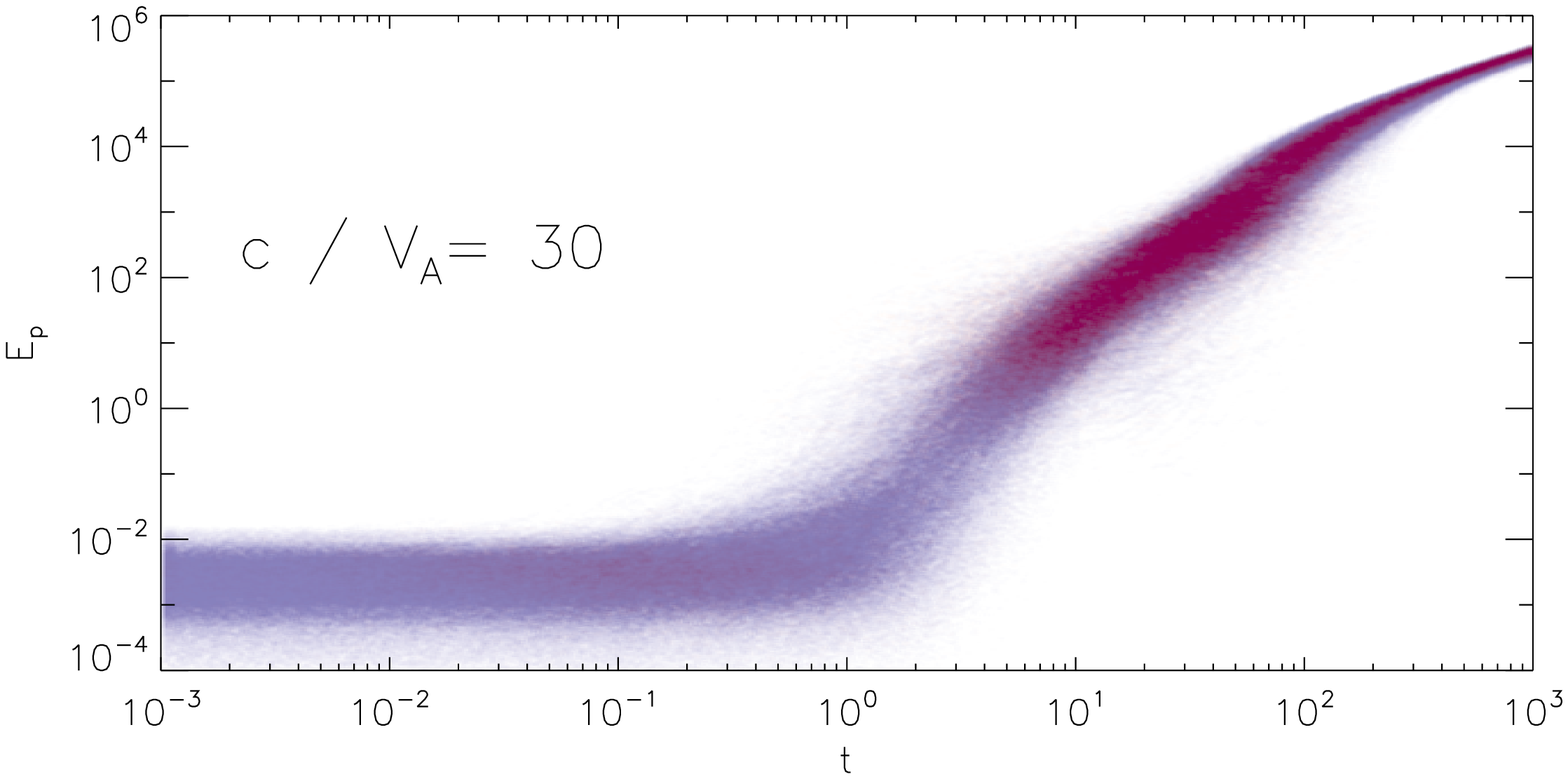}\\
\includegraphics[width=0.8\linewidth,angle=0]{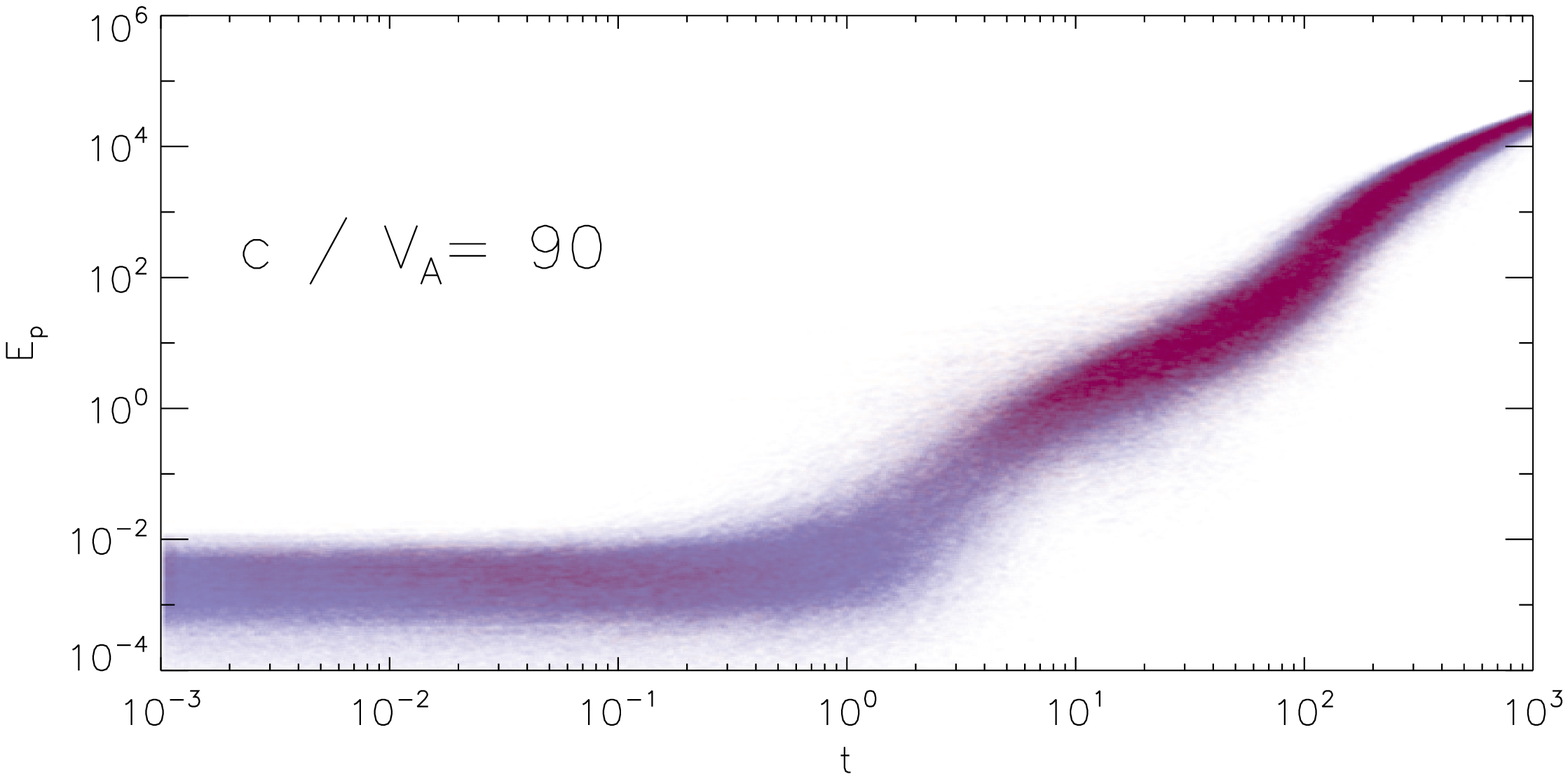}\\
 \includegraphics[width=0.8\linewidth,angle=0]{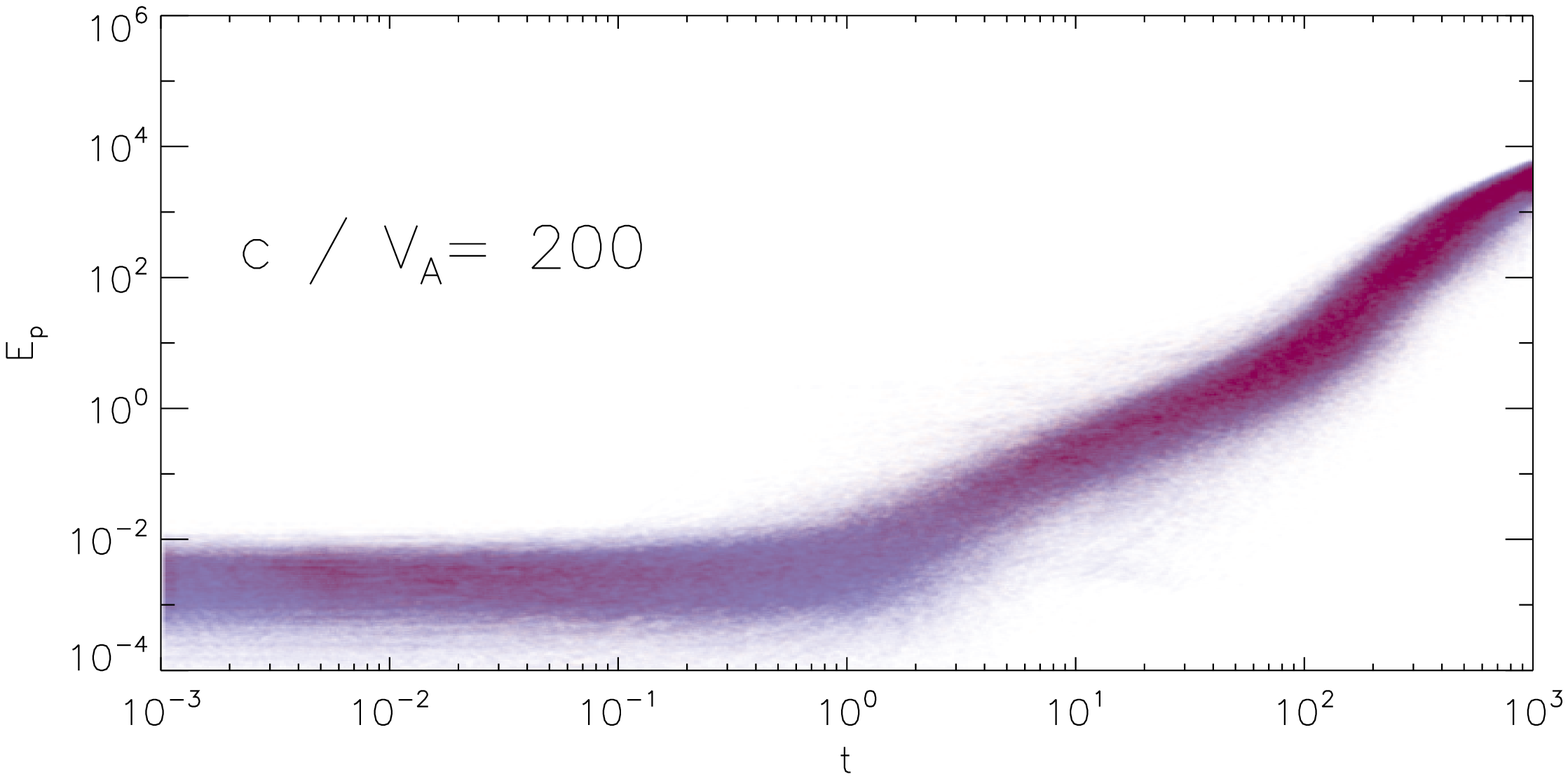}
\caption{Proton kinetic energy evolution for $c=30$, $90$ and $200$ $V_{\rm A}$ for model I. The colors {correspond to particles accelerating their} parallel (red) and perpendicular (blue) components of {velocity} {with respect to the local mean field}.}
\label{evo-maps}
\end{centering}
\end{figure}

\section{Results}
\subsection{Acceleration time}
\label{sec: time}

As mentioned above and from Eq.~(\ref{vrec}), the reconnection rate $V_{\rm
rec}$ in the presence of turbulence, for a fixed value of the anti-parallel
magnetic field component ($\equiv$ $V_{\rm A}$),  depends only on $P_{\rm inj}$
and its injection scale $k_{\rm inj}$ ($l_{\rm inj}$). The acceleration rate is
naturally expected to depend on $V_{\rm rec}$ \citep{degouveia15}. In what
follows we study the dependence of the acceleration time on {$V_{\rm A}/c$},
$P_{\rm inj}$ and $l_{\rm inj}$.

For computing the mean acceleration time as a function of energy, we calculate
$t_{\rm acc}= \sum_{i = 1}^{N_E} t_{i}/N_E$ 
where $N_{E}$ is the number of
particles with energy between $E$ and $E+{\Delta}E$, and {$t_{i}$} is the
acceleration time of particle $i$; the corresponding standard deviation is
{$\sigma^2 = \sum_{1}^{N_E}(t_{\rm acc} - {\bar{t}})^2/(N_{E}-1)$}.


\subsubsection{Dependence of the acceleration time on $V_{\rm A}/c$ and the particle energy}
\label{sec:V_A}

For fixed different values of the turbulence injection parameters we
study the particle acceleration as a function of $V_{\rm A}/c$.

In Figure
~\ref{evo-maps} we show  the kinetic energy evolution {in logarithmic diagrams}  for $c = $ $30$, $90$ and
$200$ $V_{\rm A}$ for model I.  The initial particles distribution is the same in
all cases. 
{Particles are injected with a thermal distribution  with a temperature corresponding to the sound speed of the isothermal MHD model \citep[see][]{kowal11}.} 
{The maps depict  energies for both  the parallel (red)  and the perpendicular (blue) particle velocity components\footnote{The number of  particles being accelerated preferentially in the perpendicular or parallel direction is counted considering: $u_{//} > u_{\perp}$  and ${\rm d}u_{//}/{\rm d}t > 0$ in the parallel case, and $u_{\perp} > u_{//}$  and ${\rm d}u_{\perp}/{\rm d}t > 0$ in the perpendicular one.}. }
Initially, particles suffer drift acceleration under
the effect of the magnetic field gradients, which is slow {(see the left part of each diagram up to $t =  10^0$)};  when they arrive in the current
sheet they start bouncing back and forth between the converging  magnetic
fluxes: the first order Fermi process starts operating 
\citep[{see Figure 5 in}][]{kowal11}. The diagrams in Figure
~\ref{evo-maps} clearly indicate that the particles  energy
increases very fast up to a maximum value.
\footnote{We note that the regime of faster increase in the log-log diagrams of Figure
~\ref{evo-maps}, actually reflects the exponential growth of the energy with  time for each of the thousands of accelerated particles (Figs. 4 and 5 of \citealt{kowal11} show this exponential growth of energy for each test particle more clearly). This increase, which in the log-log diagrams is provided by the collective behaviour of all particle sample - can {be  fitted by} power laws with indices > 2 \citep[see][]{kowal12a}.}
 After reaching this maximum energy, the
particles are no longer confined to the current sheet and the stochastic
mechanism saturates. This can be inferred in the energy evolution,
which exhibits  a change in the growth rate {{at around $t = 2 \times 10^2$}}. After that, particles continue to
accelerate, at a slower rate, possibly due to the drift acceleration \citep[see also][]{kowal12a}.
As we decrease  $V_{\rm A}$,  the acceleration process takes {slightly} longer time to
start, as depicted in Figure~\ref{evo-maps}. The maximum energies that particles reach
also change with $V_{\rm A}$, as it can be clearly seen in  Figure~\ref{evo-maps}. The
maximum  energies for $c =$ $30$, $90$ and $200$ $V_{\rm A}$ are approximately {around} $10^{4}$~$m_{p}c^{2}$, $6 \times 10^{3}$~$m_{p}c^{2}$ and
$10^{3}$~$m_{p}c^{2}$, respectively.

In the absence of losses, the maximum energy that particles can reach
(accelerated by the first order mechanism) depends on the size of the
acceleration region (the {effective} thickness of the {reconnection layer})  and on the value of
the magnetic field. {Indeed, in} order to be confined  the particle Larmor radius should
be smaller than the thickness of the acceleration region $l_{\rm
acc}$\footnote{We are not considering radiative loses, but for protons they may be
important under extreme conditions.}. The maximum energy is  then

\begin{equation}
E_{\rm max} \simeq e l_{\rm acc}\,B,
\label{eq:emax}
\end{equation}
 hence it  changes with $B$ (or $V_{\rm A}/c$).

{ If we take the thickness of the acceleration region $l_{\rm acc}$ as the width  of the current sheet predicted from mass conservation,  ${\Delta} \simeq L (V_{\rm rec} / V_{\rm A})$, where $L$ is the length of the reconnection layer which in our case is the size of the computational domain,  then according to Lazarian-Vishniac theory $E_{\rm max} \simeq \sqrt{4 \pi \rho} P_{\rm inj}^{1/2} l_{\rm inj} V_{\rm A}^{-1/2}$ \cite[e.g.,][]{eyink2011}. However, in the 3D simulations employed here  \citep[see also,][]{kowal12a} {the turbulence was injected in a region around the current sheet of thickness $\sim$ $0.4L$,} {therefore}, the maximum energy should be constrained by ${0.4}L$ rather than by ${\Delta}$ and in this case $E_{\rm max} \propto {0.4} L~B \propto {0.4} L~V_{\rm A}$.} This is consistent with the values of the maximum
energies inferred from Figure~1 which increase with $V_{\rm A}$.

\begin{figure}
\begin{centering}
\includegraphics[width=0.8\linewidth,angle=270 ]{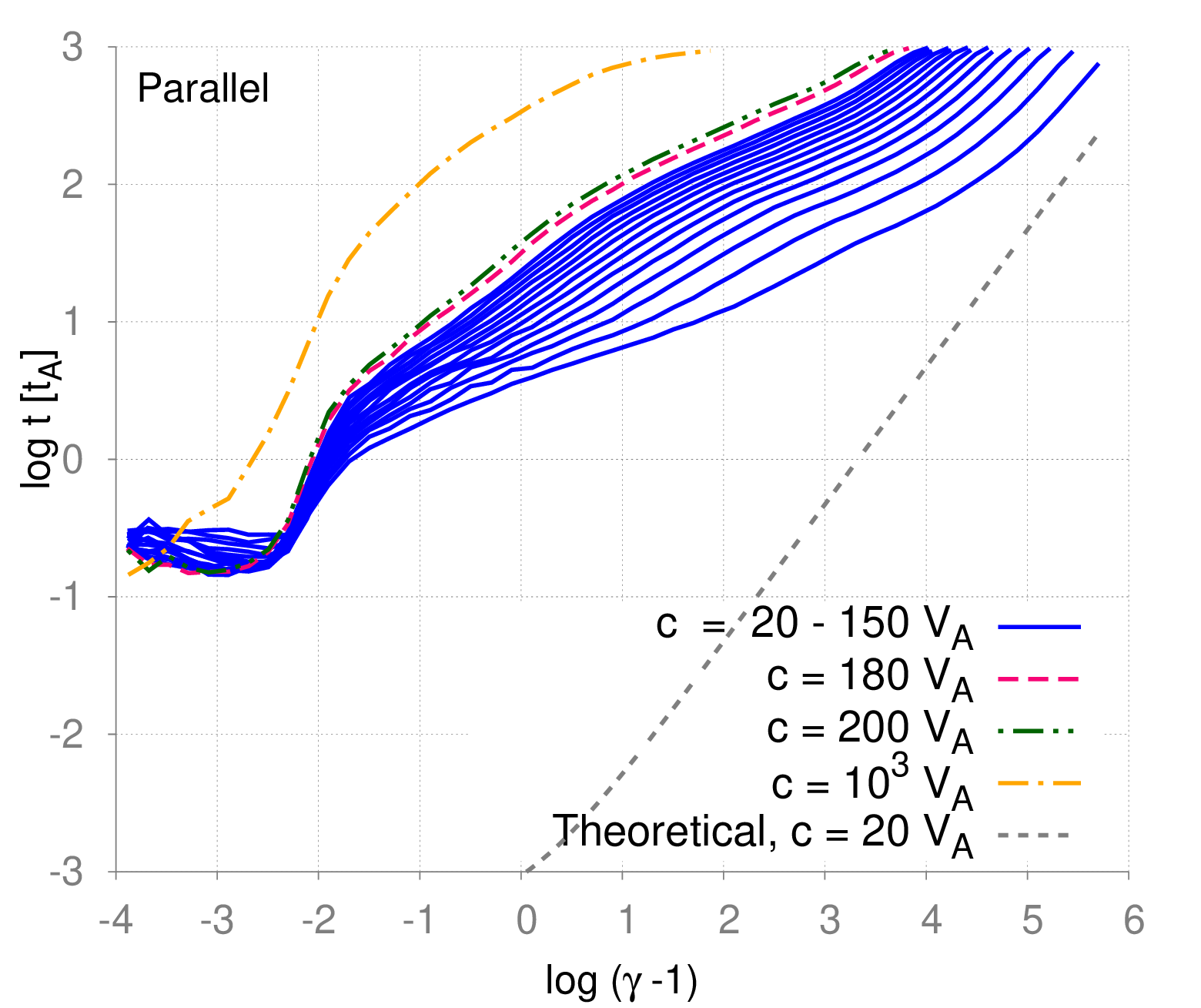}\\
\includegraphics[width=0.8\linewidth,angle=270]{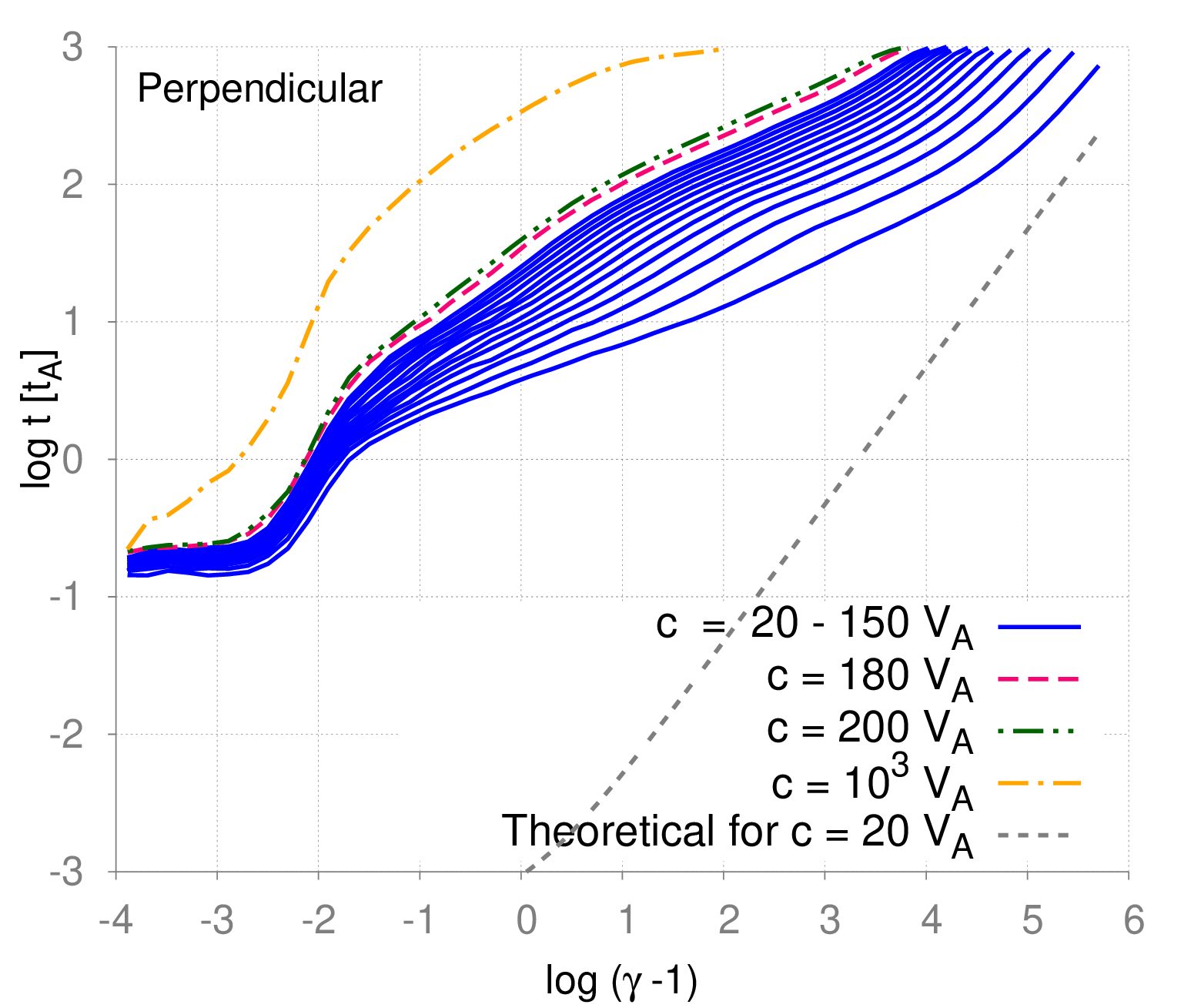}
\caption{Acceleration time as a function of particle {kinetic energy (normalized by the rest mass)} for different
values of $c/V_{\rm A}$, from 20 to 1000, for both  parallel (upper panel) and
perpendicular particle velocity components (bottom panel) {for model I}. The blue
lines correspond to values  $c / V_{\rm A}$ from 20 to 150, with $\Delta
\left( c/V_{\rm A} \right) = 10$. {The dashed grey line gives the predicted minimum
acceleration time  (Eq. \ref{eq:threshold})  $t_{\rm thr}$ for the  model with  $V_{\rm A} = c /
20$.}
}
\label{rates}
\end{centering}
\end{figure}


\begin{figure}
\includegraphics[angle=0,scale=.47]{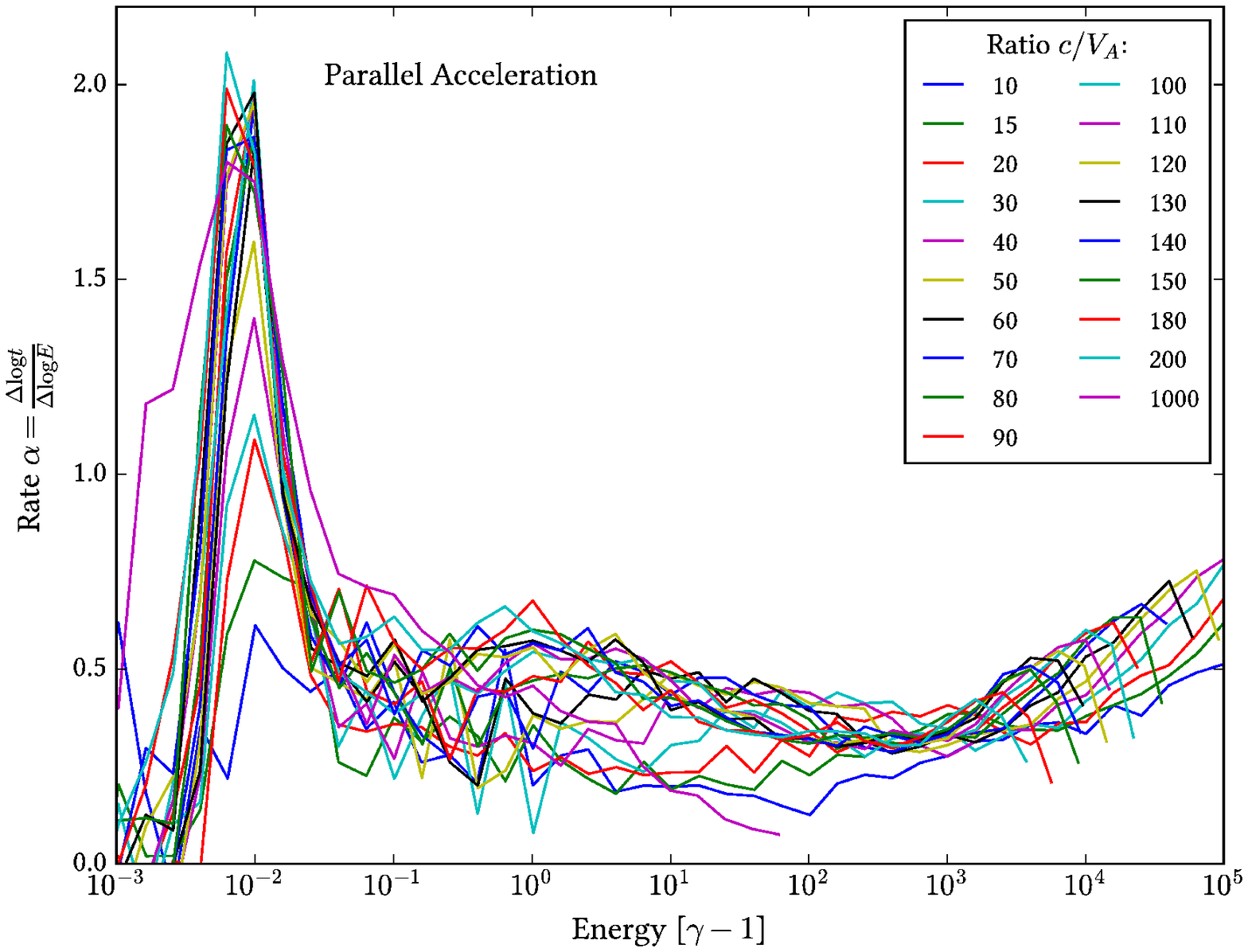}\\
\includegraphics[angle=0,scale=.47]{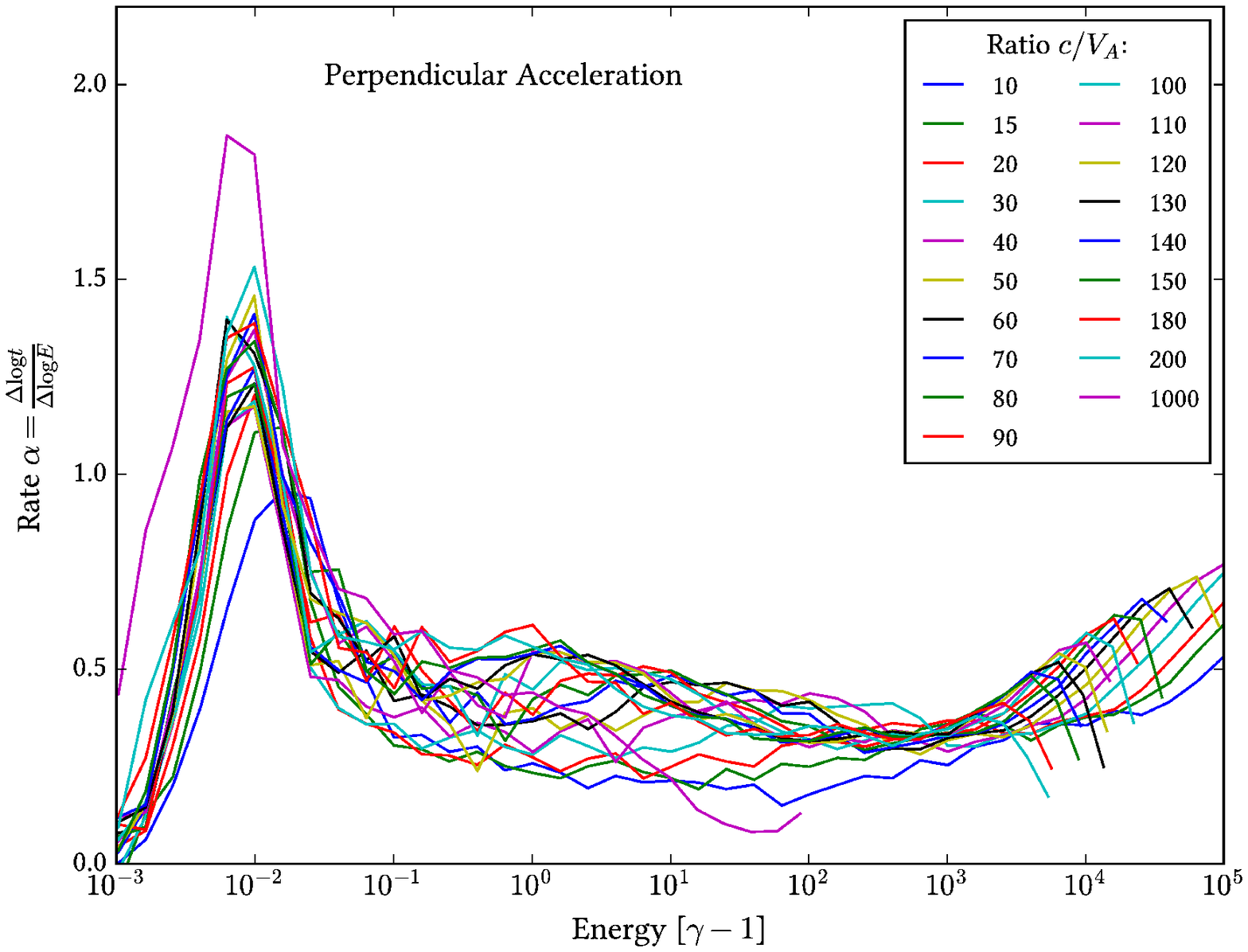}
\caption{Power-law index $\alpha$ of the acceleration time for both parallel (up) and and perpendicular acceleration (bottom) as a function of the kinetic energy.}
\label{fig:po-ekin}
\end{figure}

\begin{figure}
\includegraphics[angle=0,scale=.5]{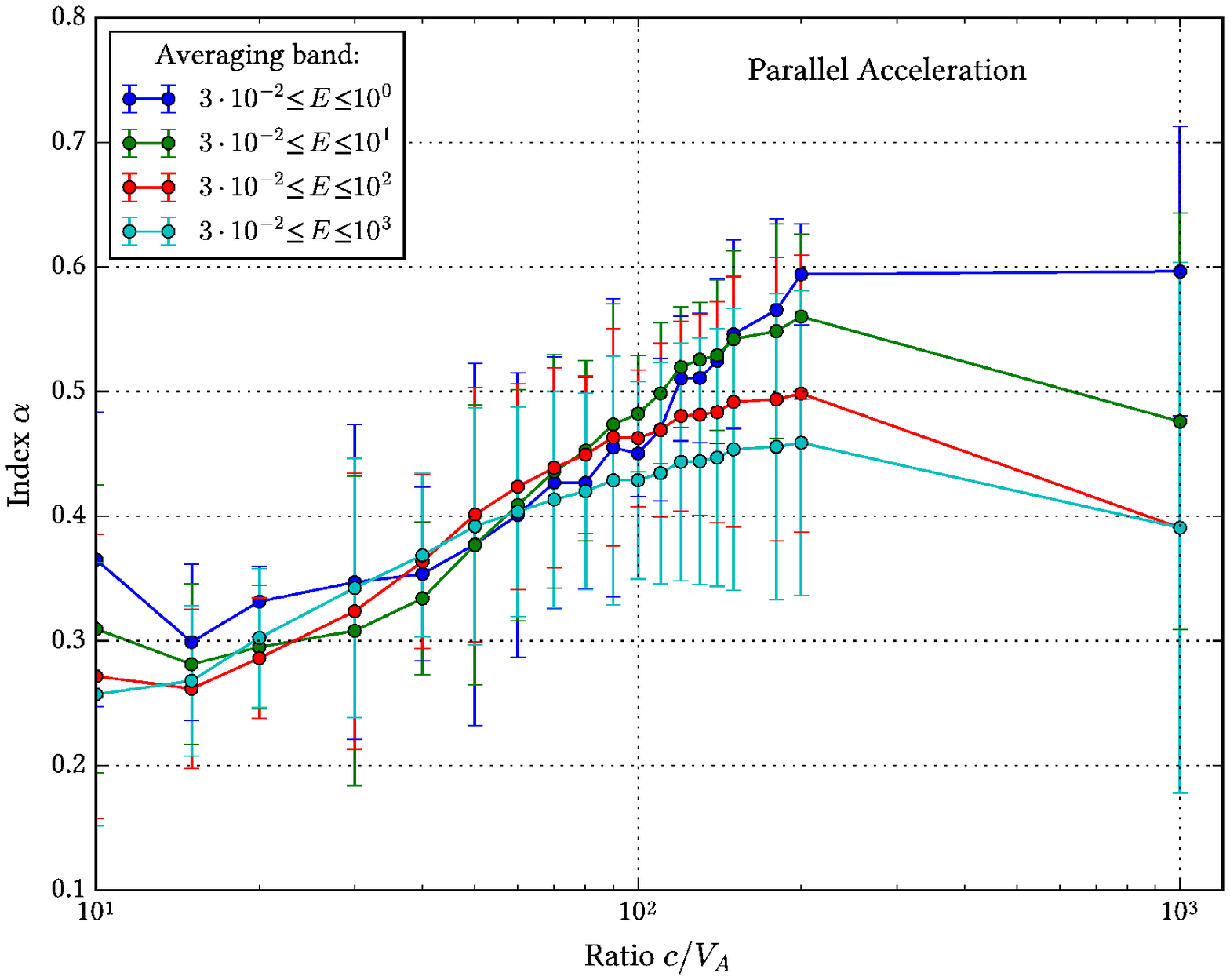}\\
\includegraphics[angle=0,scale=.5]{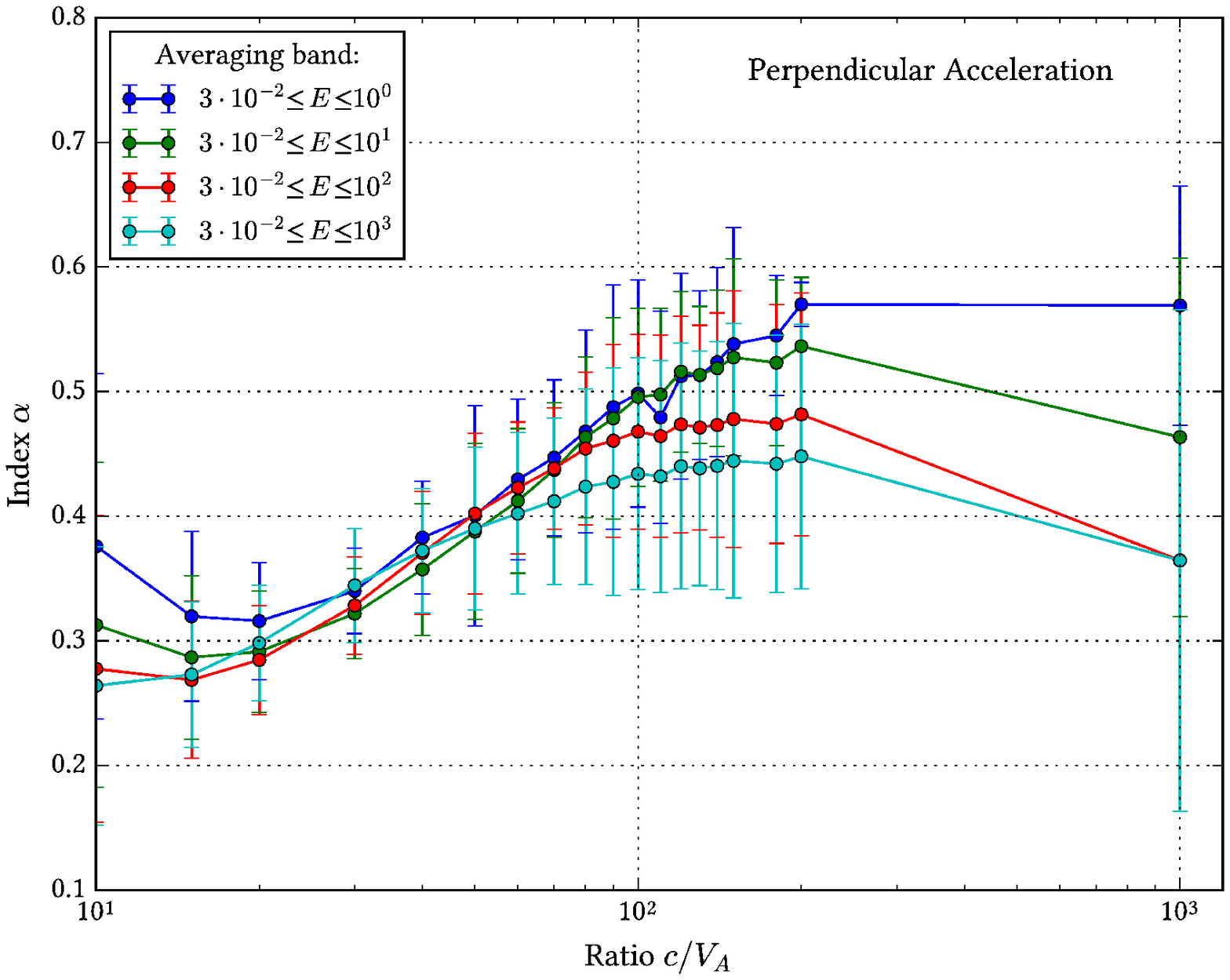}
\caption{Power-law index {$\alpha$ of the acceleration time} for both parallel (up) and and perpendicular acceleration (bottom) as a function of the  $V_{\rm A}/c$ ratio.}
\label{fig:po}
\end{figure}

\begin{figure*}
\begin{centering}
\begin{tabular}{l}
\resizebox{.27\textwidth}{!}{\includegraphics[trim= 0.cm 4.cm 0.cm 4.cm, angle=270]{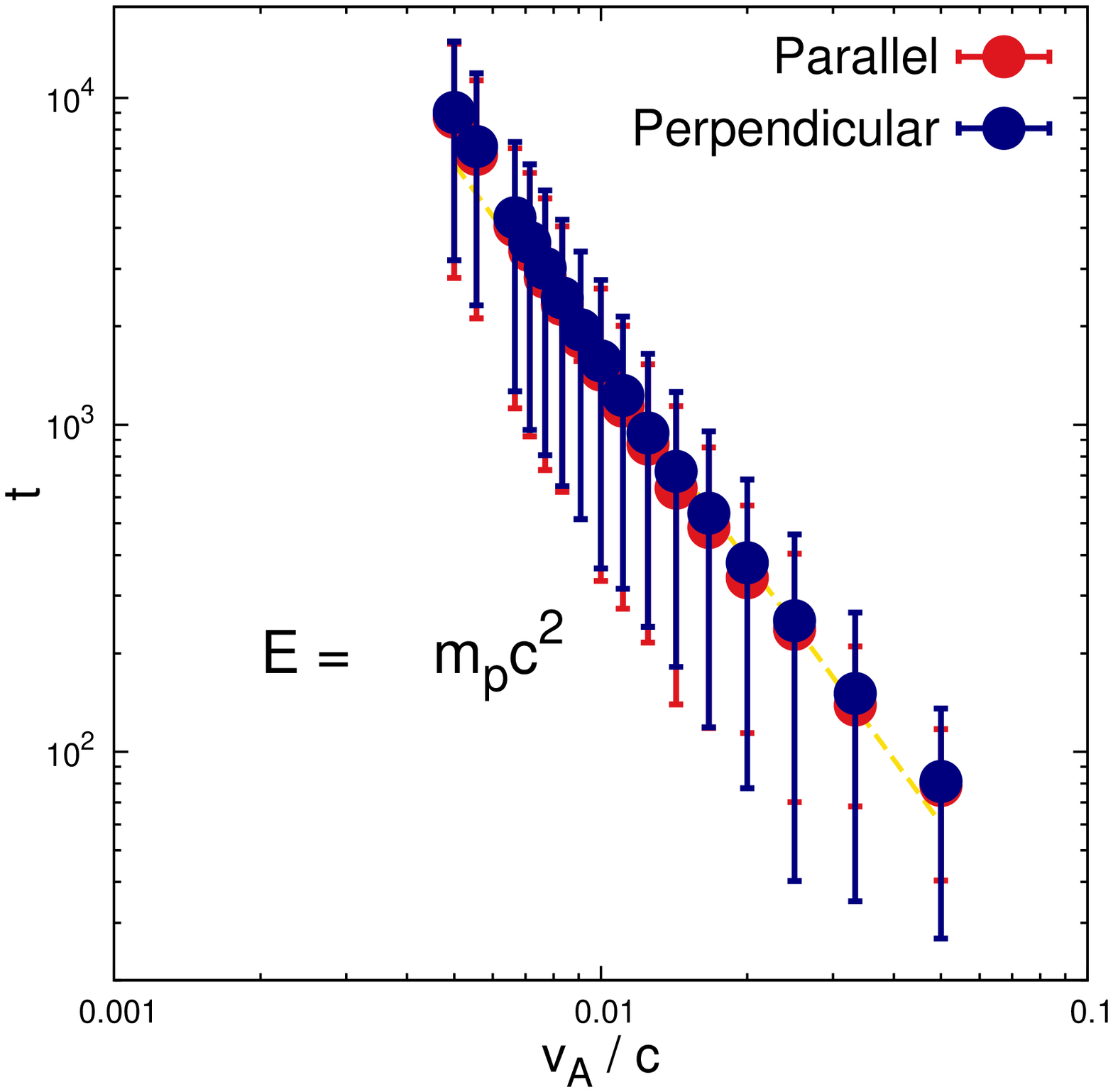}}
\resizebox{.27\textwidth}{!}{\includegraphics[trim= 0.cm 4.cm 0.cm 4.cm, angle=270]{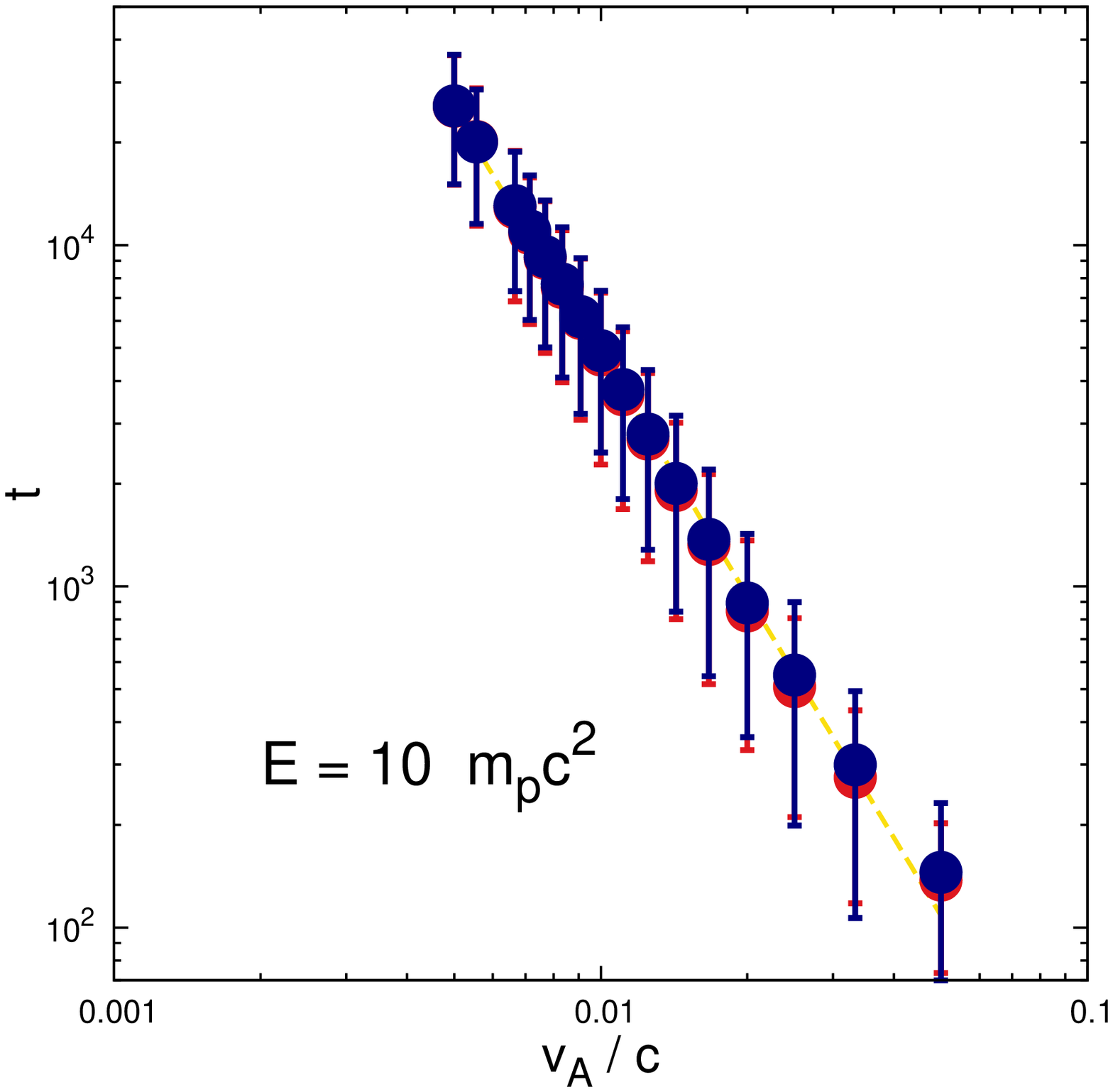}}
\resizebox{.27\textwidth}{!}{\includegraphics[trim= 0.cm 4.cm 0.cm 4.cm, angle=270]{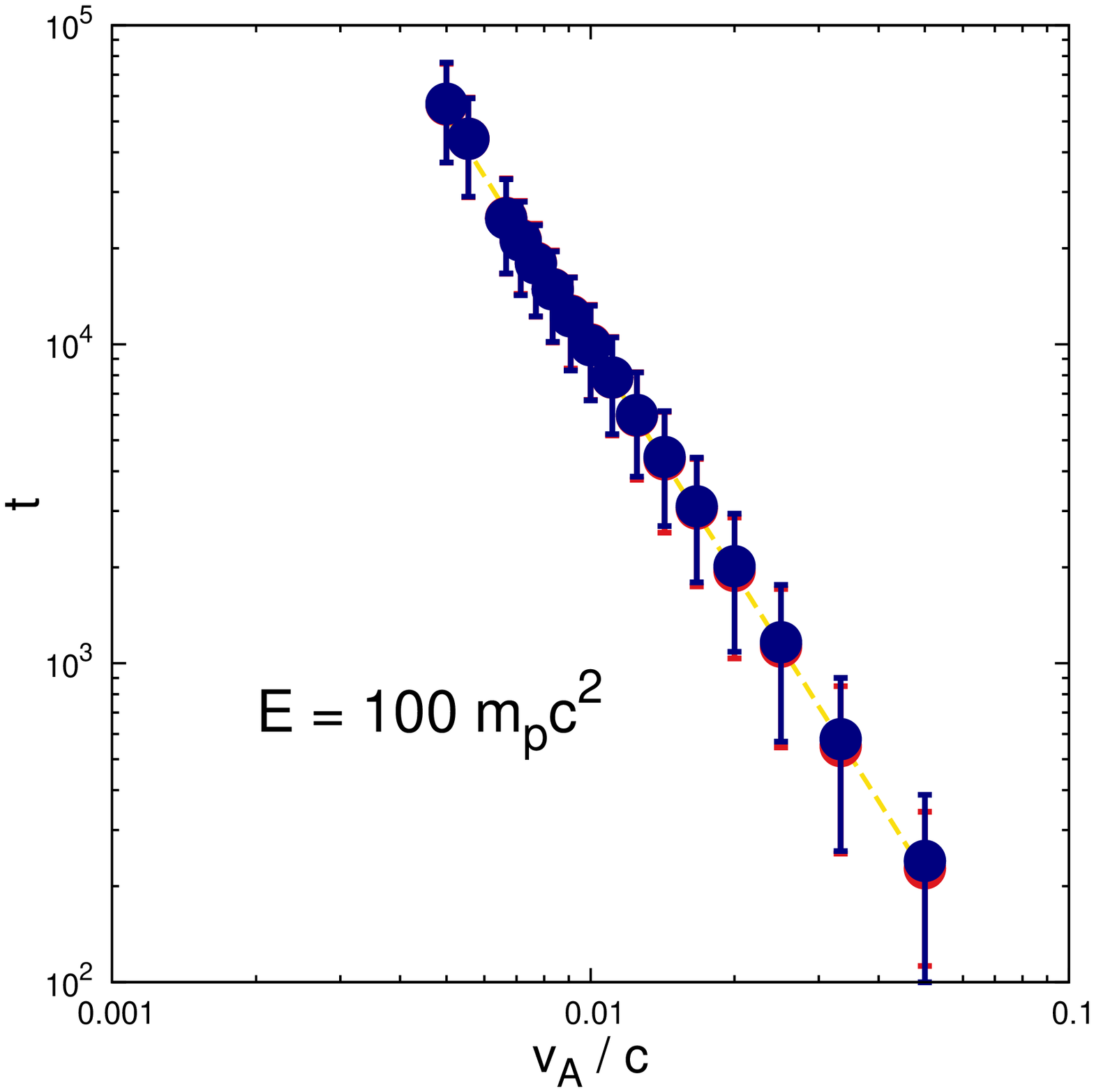}}
\end{tabular}
\caption{Mean acceleration time as a function of $V_{\rm A}/c$, for both parallel and perpendicular {particle velocity components}, for {different  proton energies}   $E = m_p c^2$, $10 m_p c^2$ and $ 10^{2} m_p c^2$. The yellow line is the best-fitted function $t_{\rm A} \propto (V_{\rm A}/c)^{-\kappa}$ (see text).}
\label{fig:t-E}
\end{centering}
\end{figure*}

Figure~\ref{rates} shows the parallel and perpendicular mean acceleration times,
as functions of {the kinetic} energy for $c = 20 - 1000 \, V_{\rm A}$, for the same MHD fast
 reconnection model (model I). The evolution of both particle velocity
components, parallel and perpendicular to the magnetic field, is tracked
separately, in order to detect anisotropies that may be important for particle
confinement, as mentioned above. The energy is in $m_p\,c^2$ units.

{ The acceleration time {in Figure 2}
can be fitted by a power law in energy, $t_{\rm acc} \propto E_p^{\alpha}$, where $\alpha$ can be estimated for a given energy band as $\alpha = \Delta \log(t_{\rm acc}) / \Delta \log(E_p)$.}

{Figure 3 presents  index $\alpha$  as a function of the kinetic energy  for the parallel and perpendicular acceleration for different values of  $c/V_A$ of model I. In both cases, there  is  a  peak in $\alpha$ around $1.5-2.0$  at very low energies ($\sim 10^{-2}$). From Figure
~\ref{evo-maps}, we see that this peak occurs when particles are undergoing  the initial slow drift acceleration {(due to the spatially varying magnetic field)} just before they enter the reconnection region {(see also Figure 5 in \citealt{kowal11}).} When particles enter the fast (exponential) acceleration regime in the reconnection zone (Figure
~\ref{evo-maps}),  $\alpha$ {drops quickly} to values below $\sim  0.6$ (at  energies $\sim {10^{-1}}$), therefore increasing the acceleration rate. In this regime, $\alpha$ remains {within the range from 0.2 to 0.6}, until the energy reaches the saturation value and  $\alpha$ starts to increase again very smoothly as particles undergo further drift acceleration,  outside the reconnection region, in agreement with Figure~\ref{evo-maps}.}

{{Figure}~\ref{fig:po} shows  the averaged value of  $\alpha$  as a
function of  $V_{\rm A}/c$ for the parallel and perpendicular components for
model I. In these diagrams, in order to evaluate $\alpha$ only in the regime of faster (first-order Fermi) acceleration, we have selected  the energy ranges for averaging $\alpha$ starting at $E_p \sim {3 \cdot 10^{-2}}$  up to $10^0$, $10^1$, $10^2$, and $10^3$. The
values of $\alpha$ span from 0.2 to 0.6. For $20 < c/V_{\rm A} < 1000$ the indices are almost
independent of the averaging energy band, as in Figure~\ref{fig:po-ekin}. For $c/V_{\rm A} < 20$, the turbulence inside the reconnection region is in the relativistic regime and the MHD simulations considered in this work are expected to be no longer valid. This explains the increase of $\alpha$ {seen in Figure~\ref{fig:po} for the smallest values of $c/V_{\rm A}$} and thus a decrease of the acceleration rate where we expected the opposite.  For $c/V_{\rm A}$ around {60 and above}, the index starts to strongly depend on the averaging energy band making  the error bars much larger, and suggesting a trend of  saturation of the value of $\alpha$ in this low $V_{\rm A}$ (and reconnection velocity) region. We further notice that the blue lines in these diagrams, which give the value of $\alpha$ averaged in the low energy band {(between 0.03 and 1)} is dominated by the region in Figure~\ref{evo-maps} right before the entering into the zone of fast} acceleration. {Since the differences of the $\alpha$ values estimated for different energy bands are still smaller than error bars}, we can conclude that the characteristic values of $\alpha$ in the first-order  Fermi  regime lie between $0.2 < \alpha <0.6$ for a large range of values of  $c/V_{\rm A}$, both for the parallel and perpendicular acceleration components.}

{Now, going back to Figure~\ref{rates}, we can estimate analytically the minimum acceleration time as} \citep[e.g.,][]{degouveia15}:
\begin{equation}
t_{\rm thr} = \frac{E}{e B V_{\rm rec}}.
\label{eq:threshold}
\end{equation}
In {Figure}~\ref{rates} we show the value of $t_{\rm thr}$ calculated
for $V_{\rm A} = c / 20$, and $V_{\rm rec}$ $\sim$ 0.1~$V_{\rm A}$. We
clearly see that the acceleration times we obtained for the model {with
$V_{\rm A} = c / 20$  are longer than this lower limit $t_{\rm thr}$ which is
attained only for  the maximum energy  in the upper right part of the diagram, as one expect since the expression above is actually valid for the maximum energy (see Eq.~\ref{eq:emax})}.

{The {above results}  are compatible with the notion that the acceleration time  $t_{\rm acc}$  must decrease with $V_{\rm rec}/c$ (see also Eq.~\ref{eq:energy_gain}). To explore more quantitatively this issue, } {we show} in Figure ~\ref{fig:t-E} the acceleration time  as a
function of $V_{\rm A}/c$ for {different}  proton {energies} $E$
$=$ $m_{p}c^2$, $10\,m_{p}c^2$ and $10^2\,m_{p}c^2$. For computing $t_{\rm acc}$
in physical units we consider {Alfv\'en time} $t_{\rm A} = L / V_{\rm A}$,
with $L = 1$, {the same time units as in the MHD simulations}. The
acceleration time $t_{\rm acc}$ decreases  with $V_{\rm A}/c$ for a fixed energy
$E$.  {Figure~\ref{fig:t-E} also shows the best fitted function  $t_{\rm acc}
(V_{\rm A}/c)$.} For { the three example} energies
considered {here} we get $t_{\rm acc} \propto (V_{\rm A}/c)^{-\kappa}$, with
$\kappa =$ $-2.09 \pm 0.06$, $-2.34 \pm 0.03$ and $-2.35 \pm 0.03$, respectively.
{The acceleration time in a DSA  process  scales with $(V_{\rm s} /
c)^{-2}$}, where $V_{\rm s}$ is the shock velocity; in the acceleration by
reconnection it seems that there is  a  stronger dependence on $V_{\rm A}/c$, at
least at high energies.


\subsubsection{Dependence {of the acceleration time} on the environment evolution}\label{time-evol}

We have  also computed the acceleration time of the particles injected in three
different dynamical time steps of the MHD turbulent reconnection site for model
I, after the turbulence  reached a statistical steady state.
The time scale of the MHD environment is {assumed to be} much longer than the particles time
scales, therefore we  expect  no significant changes {in the
particles evolution and acceleration when considering  different dynamical times
of the same MHD reconnection model} \citep{kowal12a}. From the computed
acceleration times for the {three simulated time steps of Model I,} for
both parallel and perpendicular components, we conclude that there is no
significant change as expected. We have found the same behaviour for all values
of $c/V_{\rm A}$ {({as seen in} Figure \ref{rates})}. Also, {a similar} power law $t_{\rm
acc}$ $\propto$ $E_p^{\alpha}$ with an index ${\alpha}$ $\sim$ 0.4 at relativistic
energies is observed for {these} three dynamical times.

\subsubsection{Dependence {of the acceleration time} on $P_{\rm inj}$}

The reconnection rate increases with the injection power {(see
Eq.~\ref{vrec})} and therefore,  the acceleration rate  is also expected to
increase  {for} {larger} {$P_{\rm inj}$}.  We computed the mean acceleration time for models II and
III and compared them with model I, in order to test the dependence of the
acceleration efficiency {on the turbulent}
power. Figure~\ref{fig:t-pinj} shows mean acceleration times as functions of $E$
for $c=50 \, V_{\rm A}$,  for both {velocity}
components. {
 {We clearly see that for energies smaller than the saturation value,
i.e., in the region of fast acceleration, the acceleration efficiency {increases with $P_{\rm inj}$} ({corresponding to shorter} acceleration time
$t_{\rm acc}$, as expected from Eq.~(\ref{eq:energy_gain})),
although the differences  within an order of magnitude
are, in general, encompassed by the error bars.
For  models II and III the
power-law index in the relation $t_{acc} \propto E^{\alpha}$ is a little
steeper, with $\alpha > 0.4$. At energies larger than the saturation value
(around $\gamma-1 = 10^3$ for models II and III and  $\gamma-1 > 10^5$ for model
I), {the maximum acceleration times are similar in all models}. }

{ The weak dependence seen of the acceleration time with $P_{\rm inj}$ can be attributed to the fact that particles are accelerated while being scattered by the turbulent magnetic fluctuations in the plasma ($\bf{v} \times \bf{B}$). For larger injection power  the velocity (and magnetic) fluctuations are stronger and therefore, the time the particles remain confined in  turbulent patches gaining energy may be longer than in a case with smaller injection power .}

\begin{figure}
\begin{centering}
\includegraphics[angle=0,scale=.4]{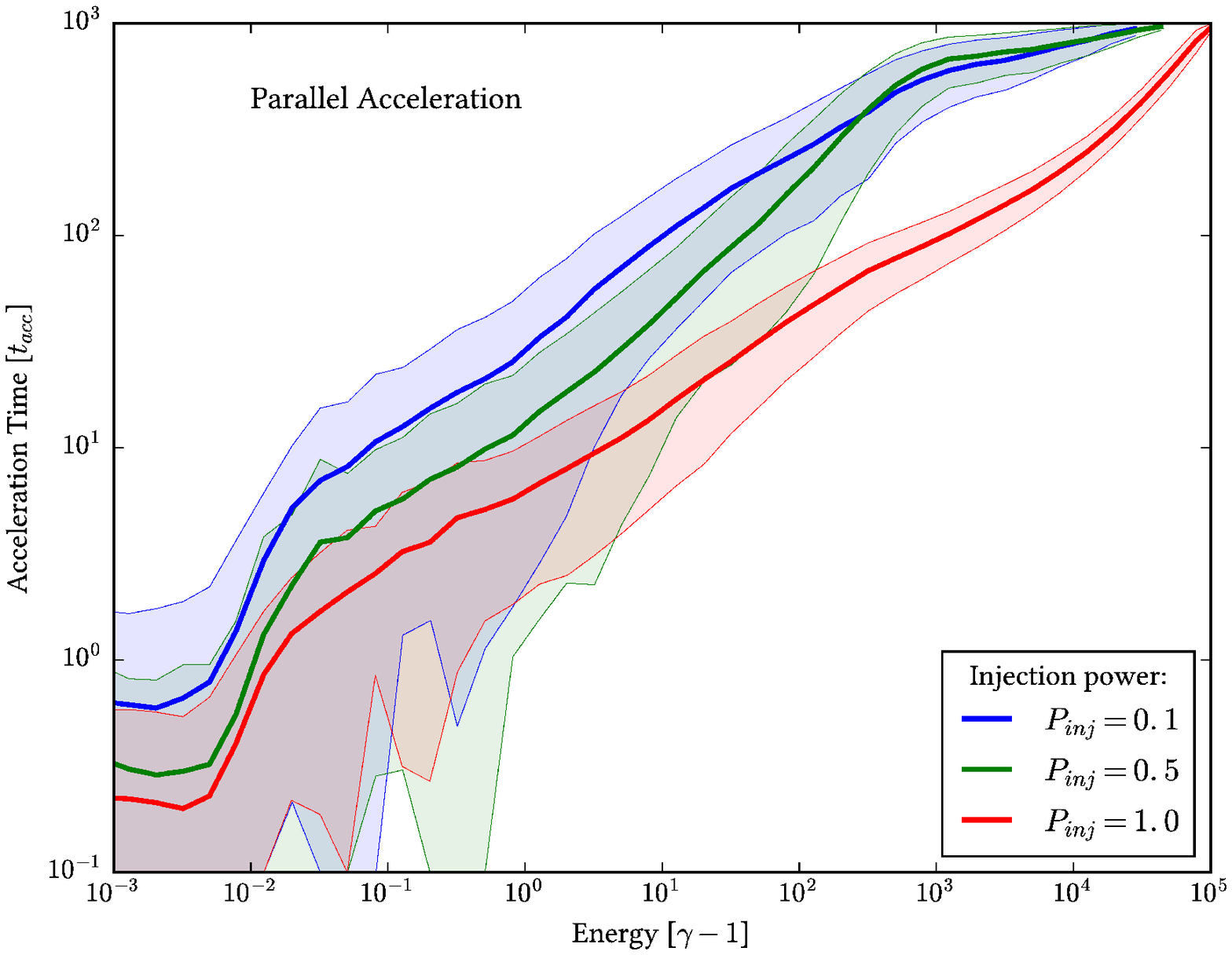}\\
\includegraphics[angle=0,scale=.4]{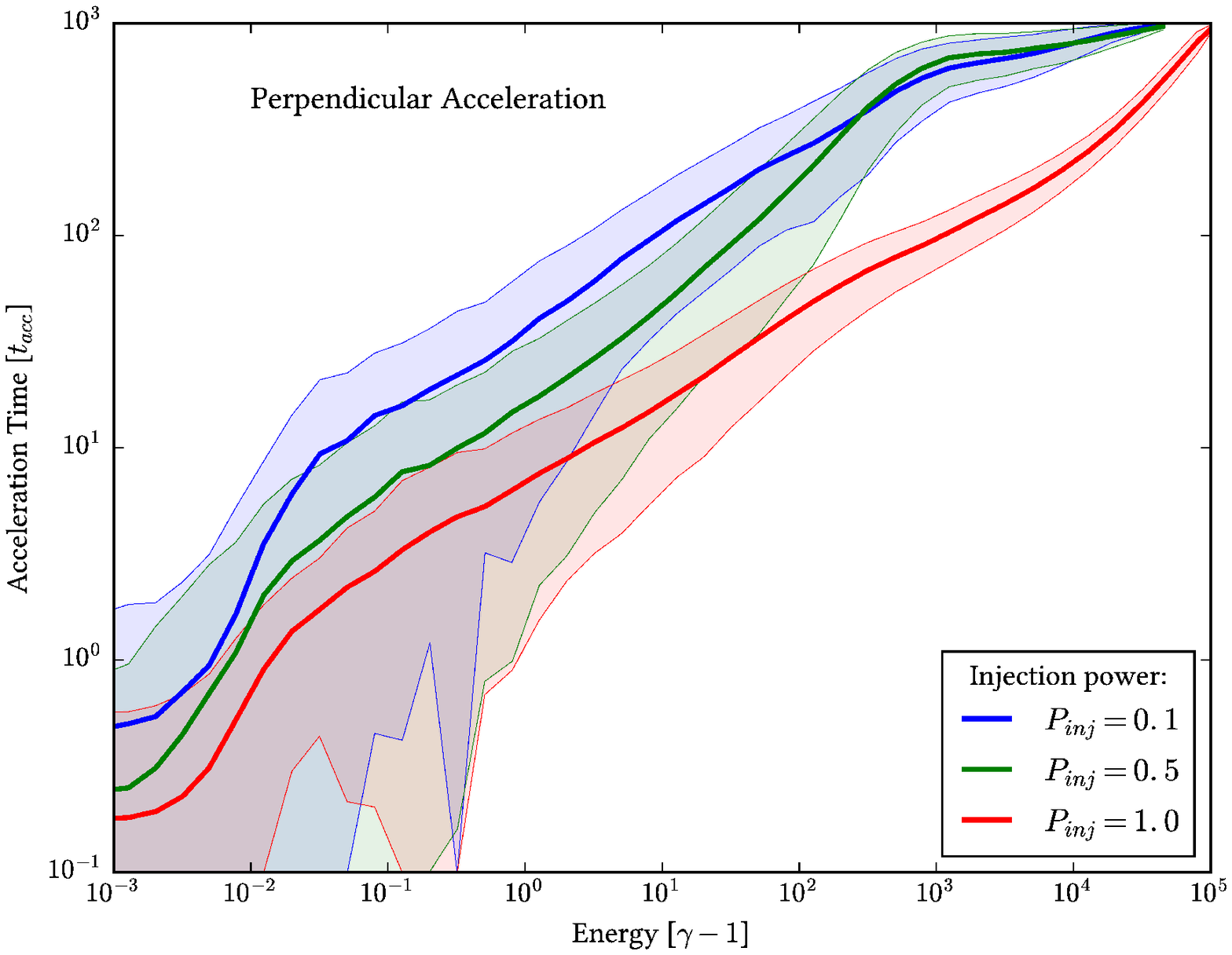}
\caption{Mean acceleration time as a function of $E$ for both parallel (top) and perpendicular (bottom) acceleration for models I, II and III {with different turbulence injection power  $P_{\rm inj}$ and the same injection scale $k_{\rm inj}=8$ (see Table~\ref{tabla}). All models have $c$ $=$ 50$V_{\rm A}$}.}
\label{fig:t-pinj}
\end{centering}
\end{figure}

\subsubsection{Dependence {of the acceleration time } on $k_{\rm inj}$}

According to Eq.~(\ref{vrec}), the reconnection rate $V_{\rm rec}$ is also
expected to increase {with injection scale of the turbulence} $l_{\rm
inj} \sim 1/k_{\rm inj}$.  We computed the mean acceleration time for models
IV and V and compared with model I to test this  dependence.
Figure~\ref{fig:t-kinj} shows the mean acceleration time as a function of $E$
for $c = 50 V_{\rm A}$, for both, parallel and perpendicular, acceleration components {with transparent areas corresponding to the estimation error of $t_{\rm acc}$. For energies smaller than $10^1$, there is no  dependence on the injection scale, which could be attributed to the fixed size of the turbulent region in all models. However, for larger energies we observed the expected behaviour, where} the efficiency of acceleration increases, or the acceleration time $t_{\rm acc}$ decreases {with the decrease of $k_{\rm inj}$}. {Although} the differences are
again within an order of magnitude at most (before the saturated energies are
reached) and generally encompassed by the error bars. For
models IV and V,  the slopes of  the relation $t_{acc} \propto E^{\alpha}$ in
the region of fast  growth  are also slightly  steeper than in model I,
with  $\alpha > 0.4$.

\begin{figure}
\begin{centering}

\includegraphics[angle=0,scale=.4]{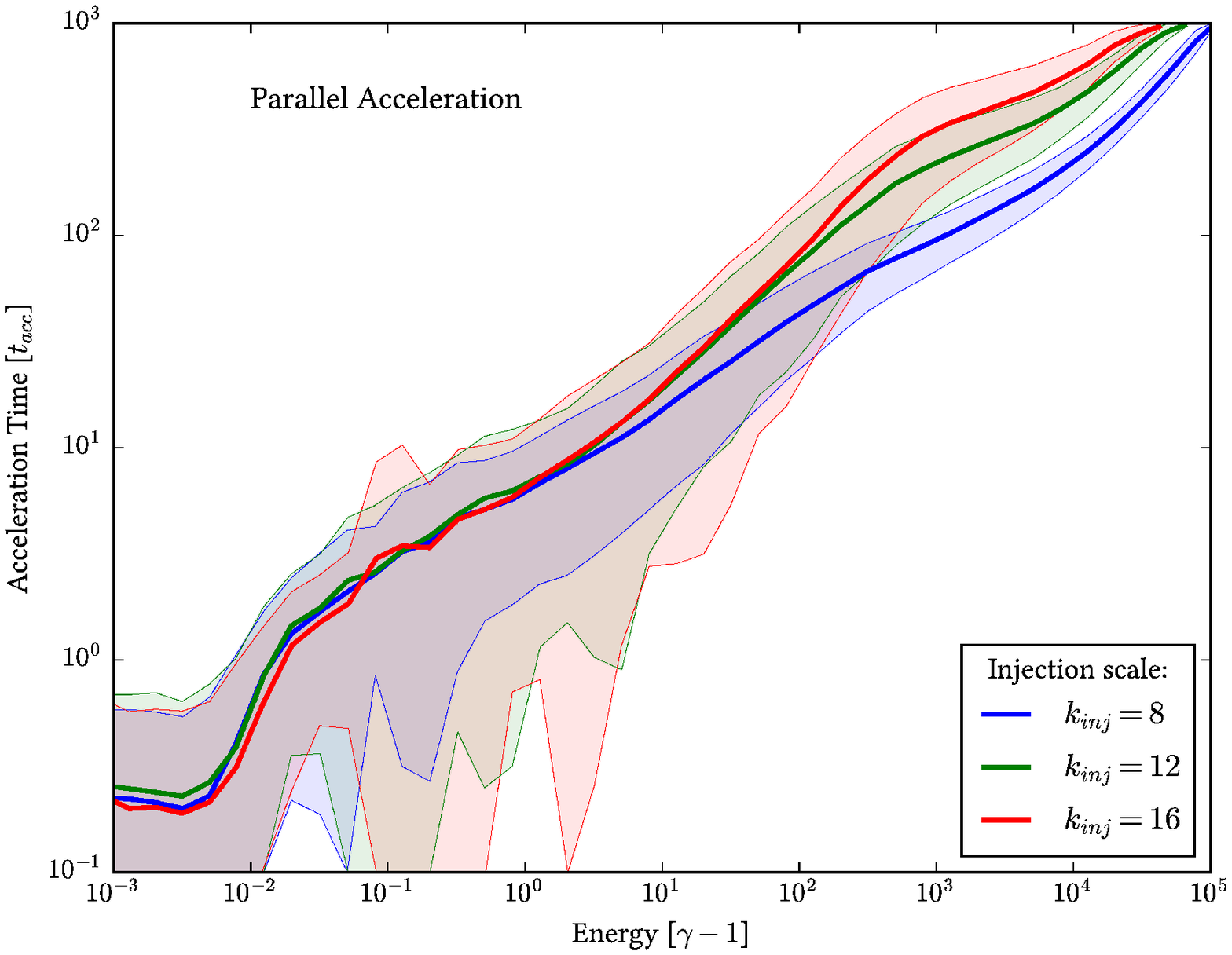}\\
\includegraphics[angle=0,scale=.4]{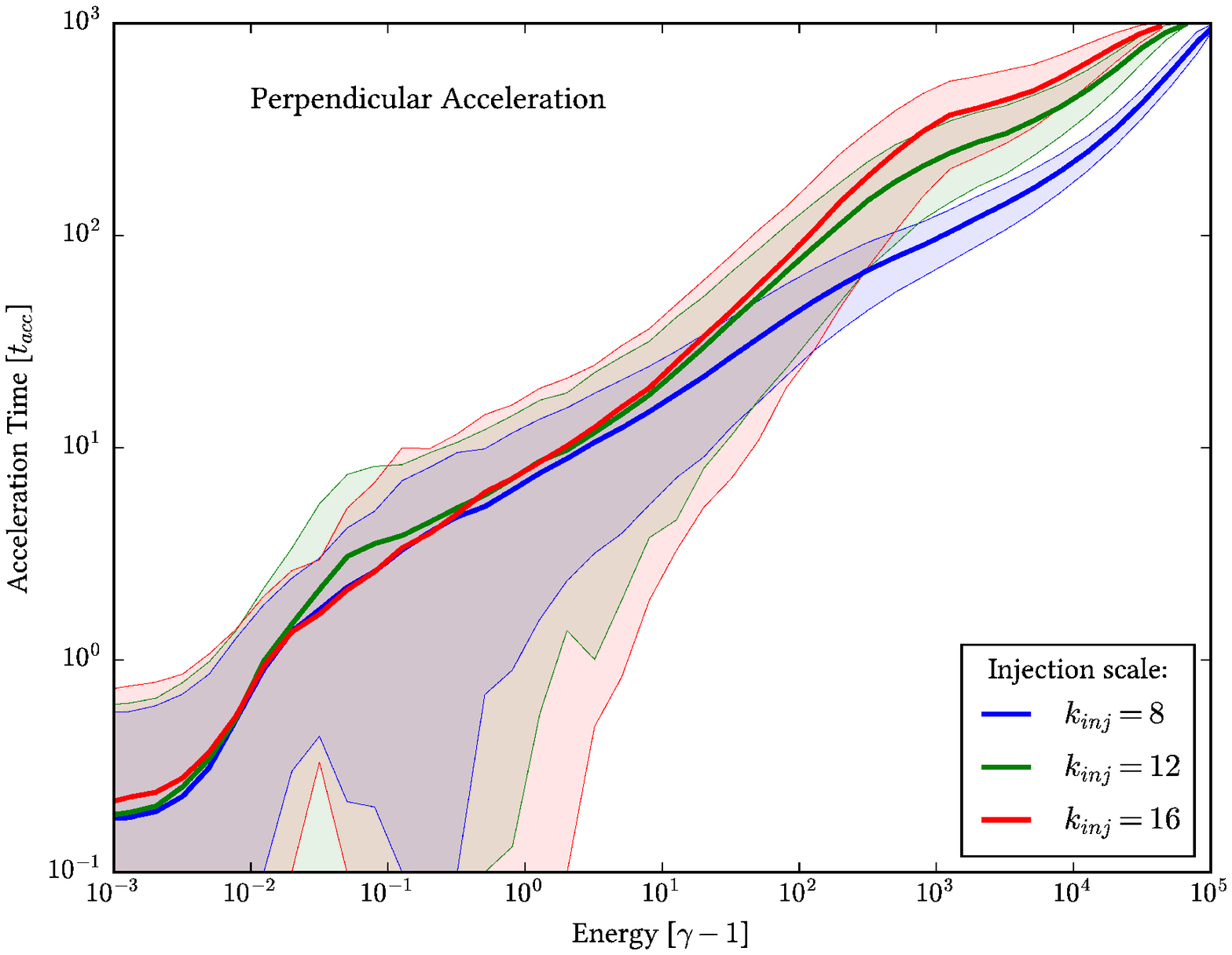}

\caption{Mean acceleration time as a function of $E$ for both parallel (top) and perpendicular (bottom) acceleration for models I, IV and V {with different turbulence injection scales  $k_{\rm inj}$  and the same injection power $P_{\rm inj}= 1$ (see Table~\ref{tabla}). All models have $c$ $=$ 50$V_{\rm A}$}.}
\label{fig:t-kinj}
\end{centering}
\end{figure}

{Comparing Figures  ~\ref{fig:t-pinj} and ~\ref{fig:t-kinj}, we note that
the acceleration time for a given $E$ has a slightly stronger dependence on
$P_{\rm inj}$ than on $k_{\rm inj}$, particularly at the highest energies near
the saturation of the fast growth.  According to Eq.~(\ref{vrec}) one might
expect the opposite, i.e., a stronger dependence on $k_{\rm inj}$.
Nevertheless, we  have also  remarked  already that the numerical simulations of
\cite{kowal09}  predict  a weaker dependence of    ${V_{\rm rec}}/{{V_{\rm
A}}}$  on $k_{\rm inj}$ than that predicted in Eq.~(\ref{vrec}) and thus we might  expect
that  this would reflect in the results for the acceleration time as well. }


\subsection{Distribution of the accelerated particles}
\label{sec:acc_dist}

\begin{figure}
 \includegraphics[width=0.7\linewidth,angle=270 ]{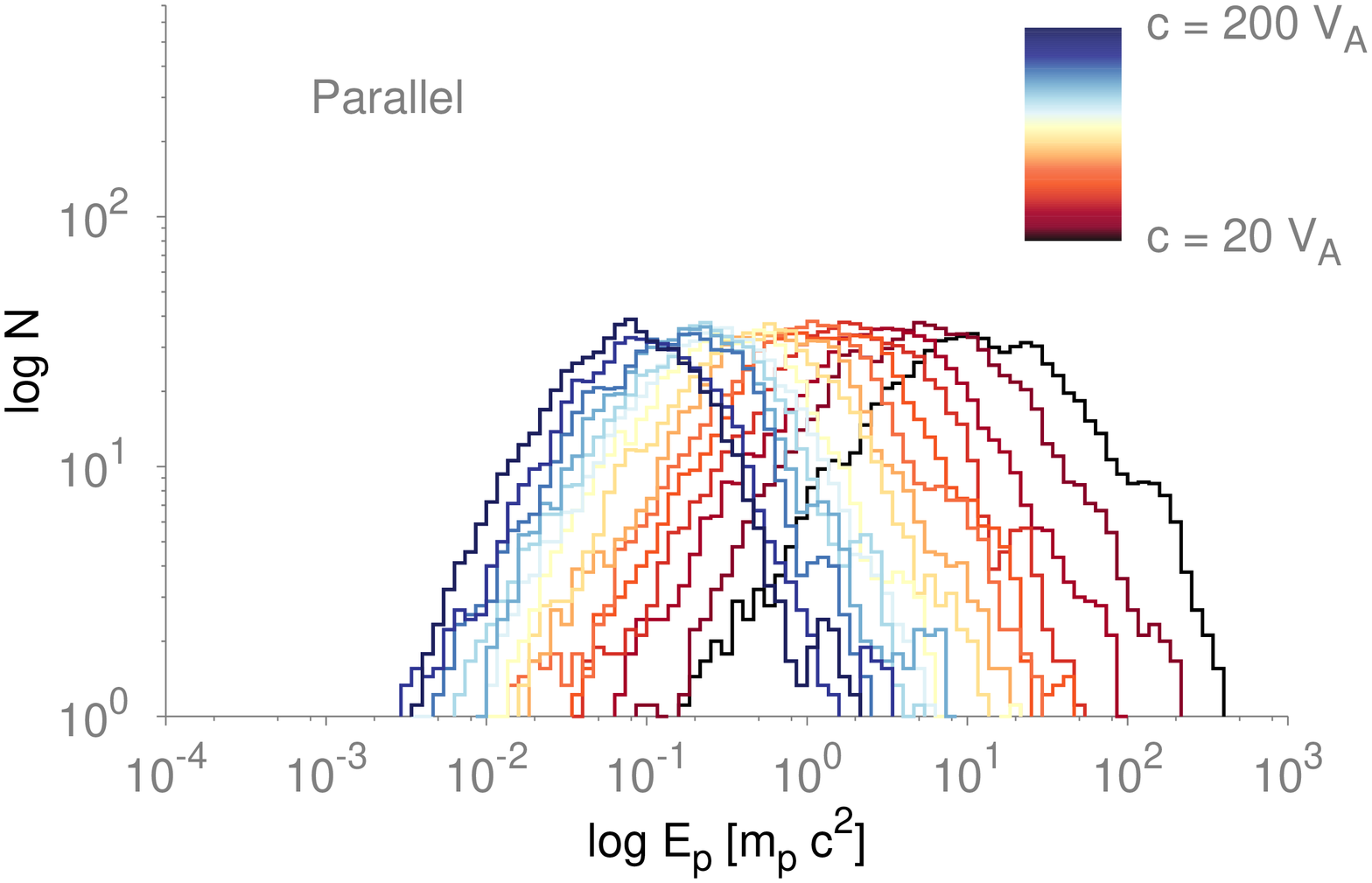}\\
 \includegraphics[width=0.7\linewidth,angle=270]{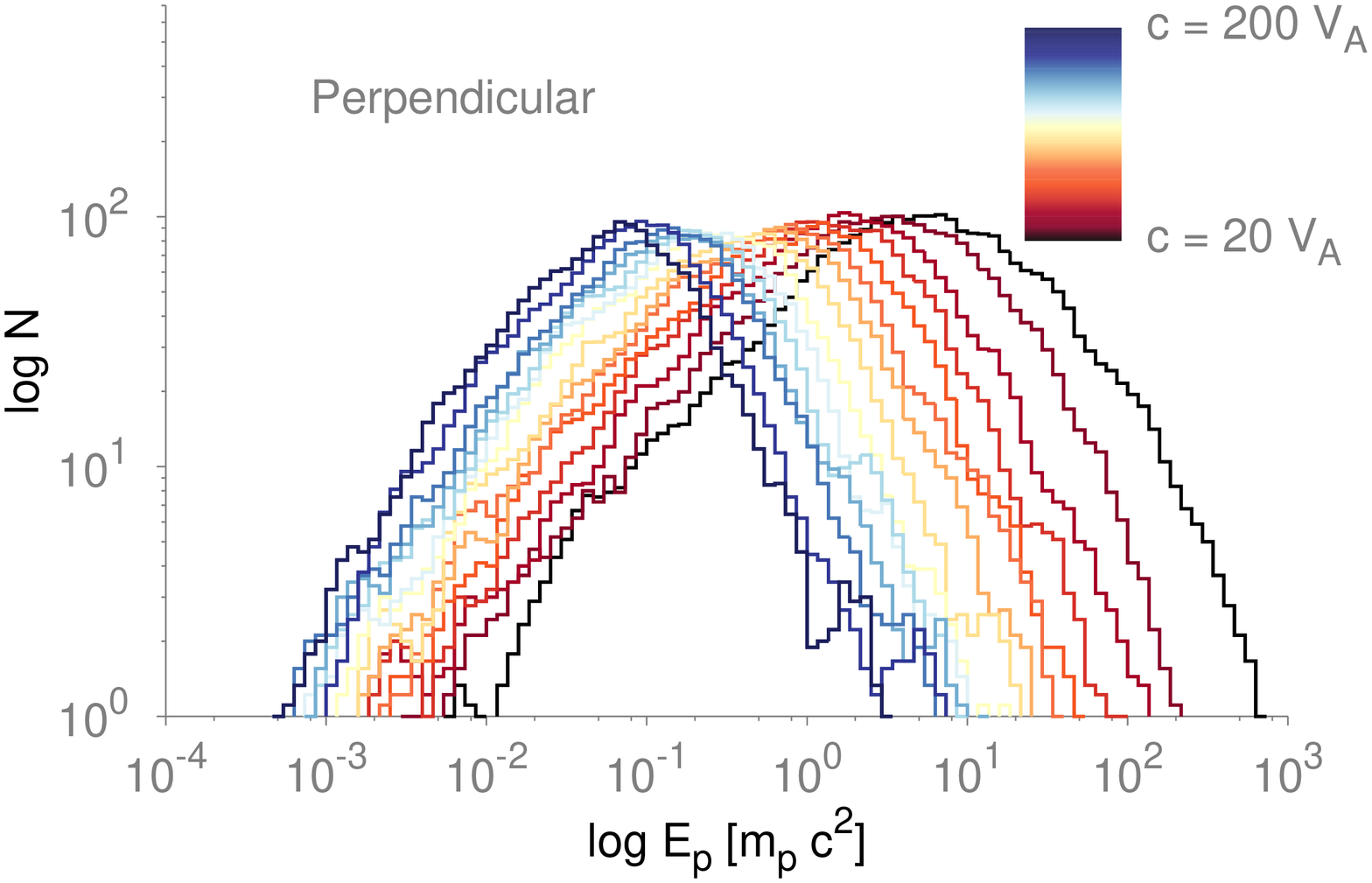}
 \caption{{Distribution} of the number of particles as a function of the particle energy in units of $m_{\rm p}c^2$ for different values of $c / V_{\rm A}$ from 20 to 200, for both the  parallel (left panel) and the perpendicular velocity (right panel) components. $t$ is 0.07 $t_{\rm A}$ after the injection, for all  $c / V_{\rm A}$ models (the other parameters correspond to those of  Model I class in Table~\ref{tabla}.}
\label{n-all}
\end{figure}

The protons initially have a thermal distribution, with {the same thermal speed for all models}. The particles begin to gain energy slowly until the Fermi mechanism starts operating. In Figure~\ref{n-all} the distribution of particles at   $t=$0.07~$t_{\rm A}$ after the injection is shown for the parallel and perpendicular components of {velocity}; the color lines represent  different values of $c / V_{\rm A}$ (from $20$ to $200$); only accelerated particles are plotted. The initial distribution is the same for all values of $V_{\rm A}/c$. {The number of particles accelerated in the direction perpendicular to the local magnetic field is larger}  than that of the particles accelerated {in the  parallel direction}. For the largest $V_{\rm A}$  higher energies {have already been reached } and the  distributions are wider in energy. In the most efficient acceleration  case  (the highest $V_{\rm A} / c$) the distribution already exhibits a power-law tail due to the Fermi acceleration \citep[see][]{kowal11, kowal12a}.

In Figure ~\ref{N-evol} the evolution of the proton distribution  during the acceleration is shown, for the parallel (red lines) and perpendicular {velocity components (blue lines) to the local magnetic field.} 
The seven time steps exhibited in the figure  correspond to $t = 10^{-0.6},\,10^{0},\,10^{0.49},\,10^{0.89},\,10^{1.97}10^{2.5},\,10^{3}$~$t_{\rm A}$. We show the distributions for the $ c =$ 30, 90 and 200~$V_{\rm A}$ {cases of model I}.  The evolution of the particle  distribution  is similar  for all values of $c / V_{\rm A}$. Initially, the particles  have a normal distribution (not shown here) and, as particles gain energy, they start to populate higher values of energy. At some point (see the second curve from left to  right in each diagram) the distribution becomes {flatter than a} normal distribution at  higher energies. Due to our numerical setup, particles  never stop accelerating {as they are continuously reinjected into the system and therefore, {the distribution shifts to higher and higher energies}. As remarked, even after  the particles  attain the maximum energies allowed in the reconnection region by the first-order Fermi process, they suffer further  drift acceleration at a smaller rate.} Thus,  in the absence of radiative  losses or a escape from the acceleration region, the particles may gain energy continuously. Most of the particles reach very high energies as time evolves.
Their distribution may even get a positive power-law index in very evolved times (as we can see in the figure), but of course in real systems we should  expect an interruption of this acceleration process with the {escape} of the particles from the finite volume of the acceleration zone and/or due to radiative losses. 

{It is interesting to note that} for all  $V_{\rm A} / c$ cases 
{shown in Figure ~\ref{N-evol}} there are more particles being accelerated in the  perpendicular than in the parallel direction to the local magnetic field. This anisotropy ensures the success of the acceleration process \citep[see discussion in][and references there in]{degouveia15}.
The effective electric field accelerating a particle (with  velocity \textbf{u}) is given by Eq.~(\ref{eq:mov}) and may {lead to} both parallel and normal  acceleration directions to the local mean field depending on the original direction of the interacting magnetic fluctuation of the turbulent flow (\textbf{v}), or in other words on the resulting direction of \textbf{(u-v)} with respect to the local \textbf{B} (given by the sum of the mean plus the fluctuating component). 
{Nevertheless, the fact that we see in Figure~\ref{N-evol} more particles in the perpendicular direction than in the parallel direction is an indication that the effective electric field  is predominant in the normal directions on average.}

During the total integrated  time considered in {Figure~\ref{N-evol}}, $t_{\rm f} = 10^3$~$t_{\rm A}$, the  particles in models with larger $V_{\rm A} / c$ reach higher energies. {Since the size of the system }$L$ is independent of $V_{\rm A}$ and is the same for all models, naturally models with larger ratios $c / V_{\rm A}$ and thus larger acceleration rate will accelerate particles to higher energies (at the  same  time interval). We should point out that {in the cases shown in this Figure,} the maximum energy achievable by the first-order mechanism (at the saturation of the fast growth) has already been reached at $t = t_{\rm f}$  (see Figure~\ref{evo-maps}).

\begin{figure*}
\begin{center}
\includegraphics[width=0.28\textwidth,angle=270 ]{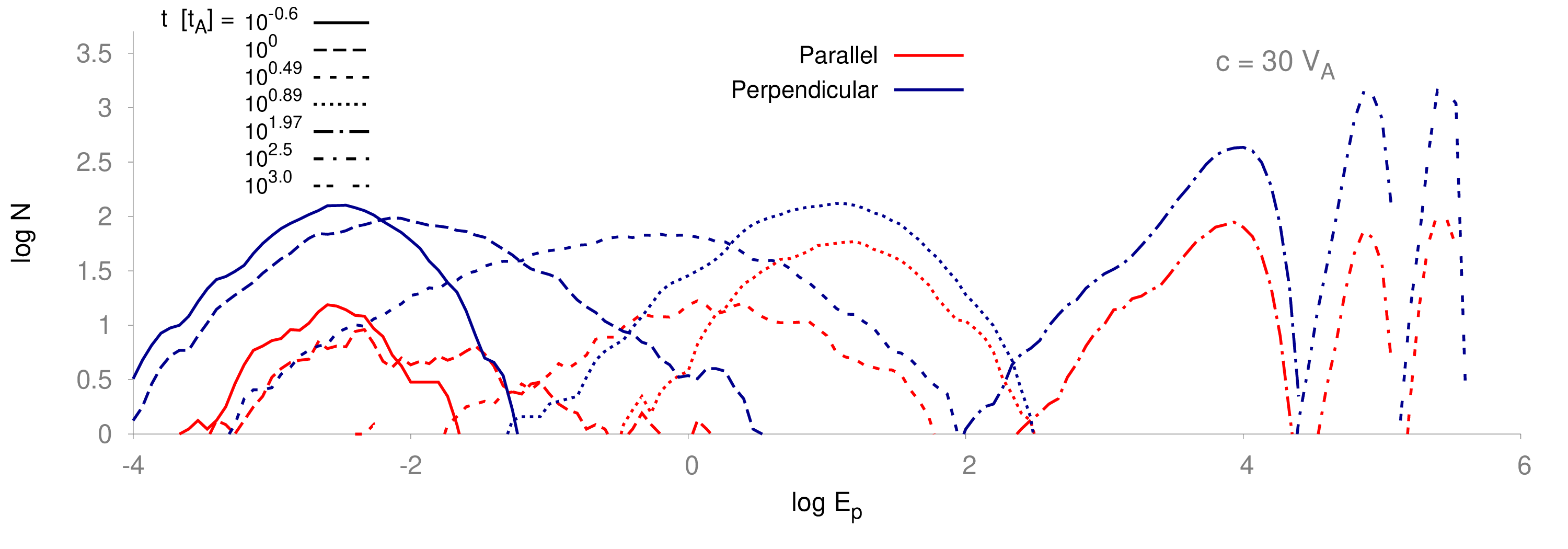}
\includegraphics[width=0.28\textwidth,angle=270 ]{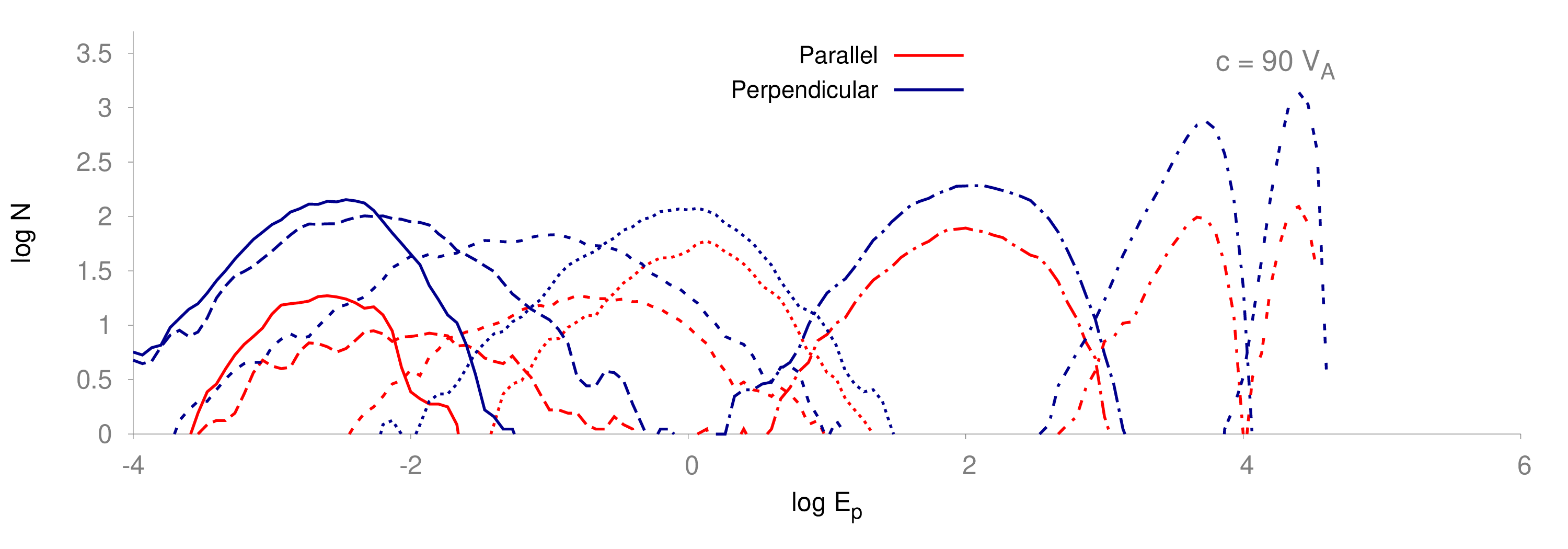}
\includegraphics[width=0.28\textwidth,angle=270 ]{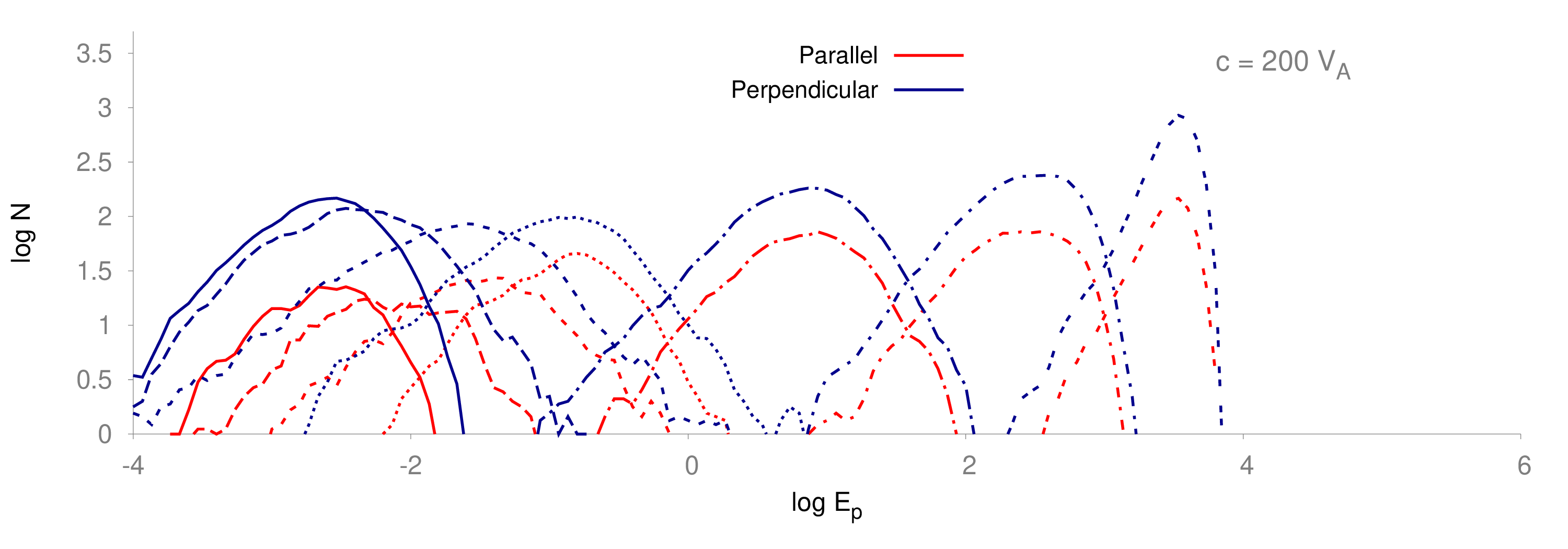}
\caption{Evolution of the number of accelerated particles  (both for the parallel (red lines) and perpendicular (blue lines) {velocity components} for $t = 10^{-3}t_{\rm A}$ to $t = 10^{3}t_{\rm A}$. We show the cases $c / V_{\rm A} =$ 30 {(top), 90 (middle) and 200 (bottom)} {for model I of Table~\ref{tabla}}. The energy is in units of $m_{\rm p}c^2$. The seven time steps {exhibited from left to right} correspond to $t = 10^{-0.6},\,10^{0},\,10^{0.49},\,10^{0.89},\,10^{1.97}10^{2.5}$ and $10^{3}$~$t_{\rm A}$.}
\label{N-evol}
\end{center}
\end{figure*}

\begin{figure}
\begin{center}
\includegraphics[width=0.3\textwidth,angle=270 ]{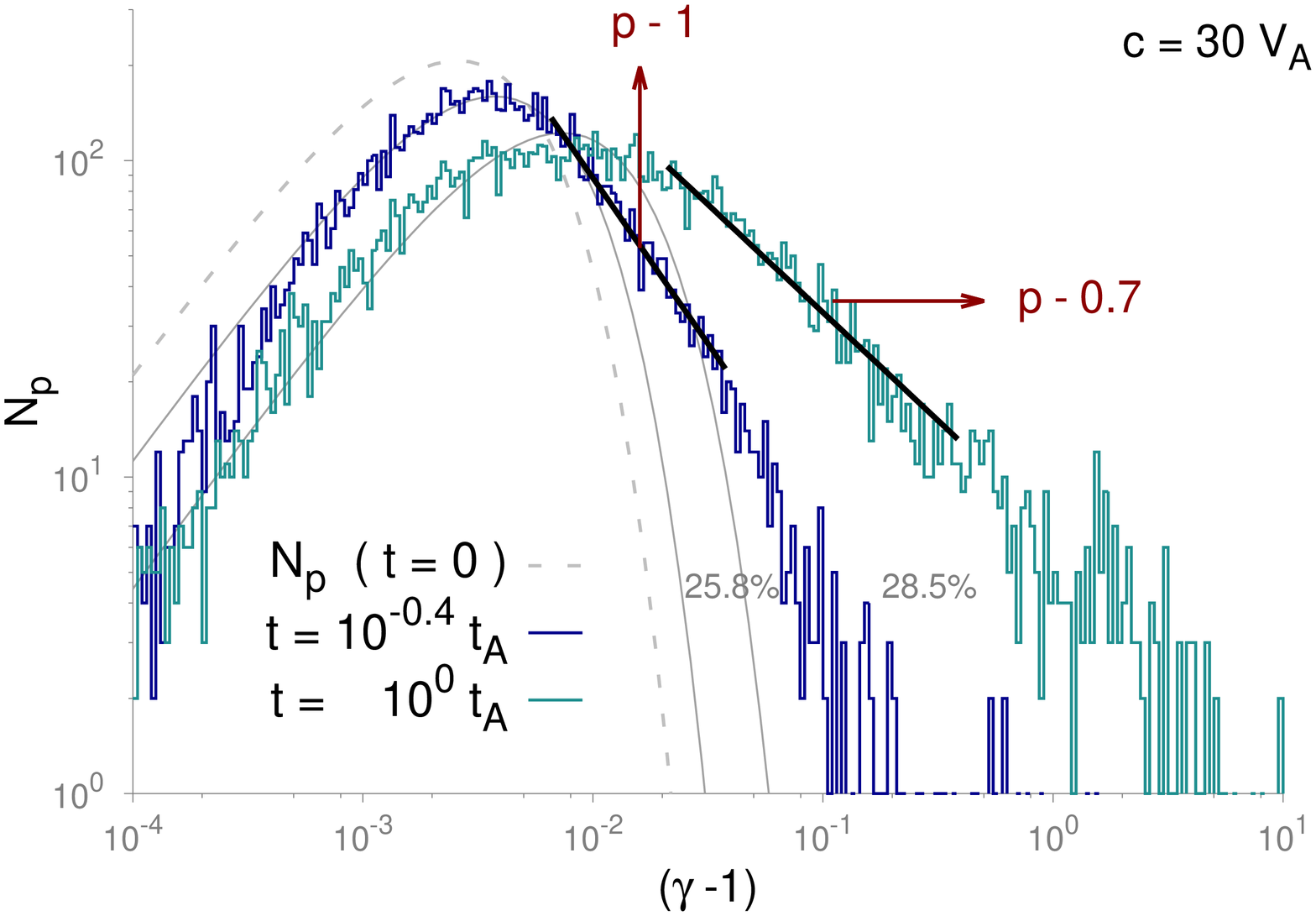}\\
\includegraphics[width=0.3\textwidth,angle=270 ]{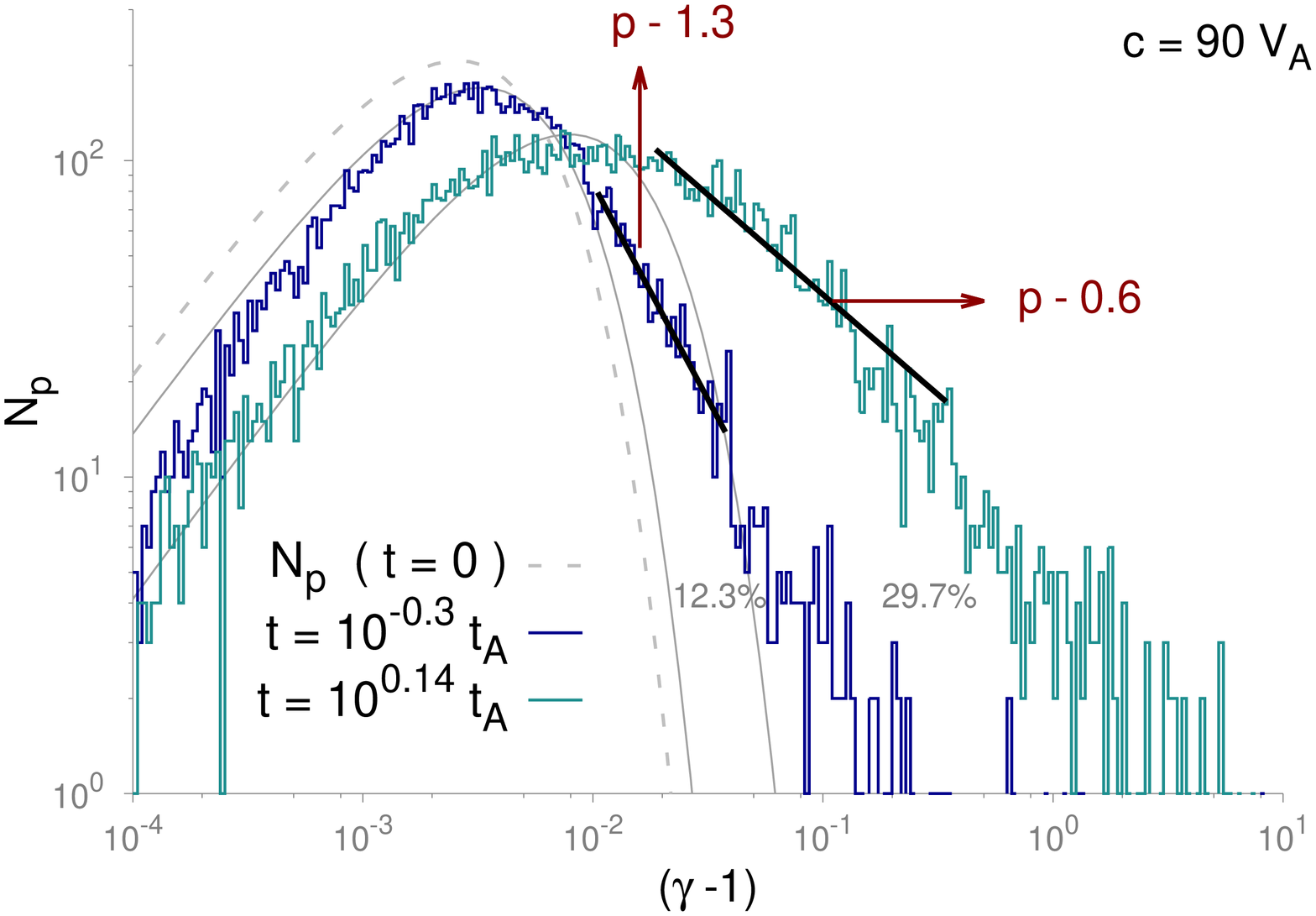}\\
\includegraphics[width=0.3\textwidth,angle=270 ]{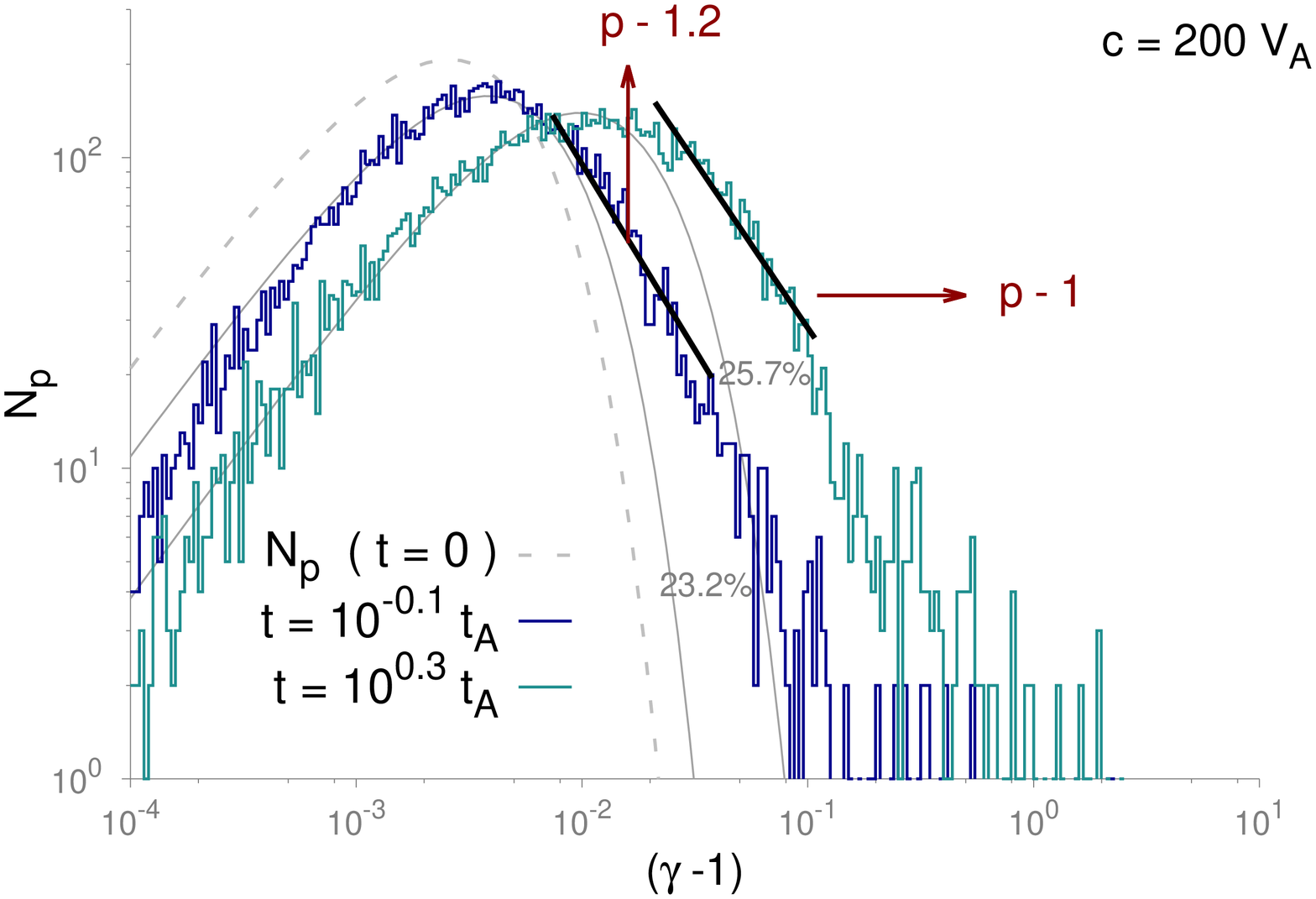}
\caption{Total particle distribution  as a function of the normalized energy at two  different timesteps for (from top to bottom) the $c / V_{\rm A}$ = 30, 90 and 200 {cases of Model I}. The dotted gray line is the initial normal distribution, which is the same for all cases. Each distribution can be  fitted by a normal distribution (shown in solid gray line) plus a power-law distribution.}
\label{fig}
\end{center}
\end{figure}

As stressed above, since in our simulations 
the particles are continuously accelerated {and there is no physical mechanism to allow {them to escape}, it is  not possible} to obtain  the actual distribution of the accelerated particles. Nonetheless, we can make some estimate  of the power-law index of the distribution  soon after the particles start to populate the high-energy tail.  

In  Figure ~\ref{fig} we show the total number (not only the accelerated ones) of particles as a function of energy for two different time steps. Each figure corresponds to the cases  $c / V_{\rm A} =$ 30, 90 and 200 {of Model I}. The initial normal distribution ($t=0$) is shown in  gray dashed lines. The earliest time step plotted in each case corresponds to the approximate time   when a high-energy power-law tail starts to form {(i.e., when particles reach kinetic energies larger than $\sim 10^{-2} mc^2$, according to  Figure~\ref{evo-maps})}; the second time corresponds to a little longer time step.  
In each time step we can distinguish two components in the distributions: a normal one (shown in solid gray line) + a power-law tail {that fits the high energies} starting at {a certain energy we denote as $E_{\rm pl}$}. We see that the first power-law index at the earlier times can be fitted {by  $\sim$ $p =  - 1.3$ to $-1$.} 
{However, these values would be a little {smaller} (corresponding to slightly  steeper spectra) if we had taken the spectra at a little earlier time. Therefore, these values can be taken as approximate ones.} The second power-law at later times  in all the cases is flatter due to the effects discussed above 
{and therefore, they must be taken only  as illustrative of the limitations of the method. 
Of course, in realistic systems, the presence of  physical particle {escape} from the acceleration zone, radiative losses and dynamical feedback of the accelerated particles into the plasma will result {in} steeper spectrum  in the late times too {($|p| >1$)}. 
We have also estimated the percentage of particles with energies $E > E_{\rm pl}$ (populating the high-energy region of the distribution). In all the cases this percentage increases from the first to the second time step.

\subsubsection{Dependence {of the particles spectrum} on $P_{\rm inj}$}

For all models of Table~\ref{tabla},  the particle distribution behaves in the same way
{as in model I above}; as particles reach higher energies {they
start to develop a power-law. } The {anisotropic} distributions  with the
production of  more accelerated particles in the perpendicular  than in the
parallel direction  are also present  in models II and III (for the studied
cases  $c / V_{\rm A} =$ 50, 100, and 150).

A considerable dependence on $P_{\rm inj}$ is found in the evolution of the
distribution of the accelerated particles.  At each given  time, model I which
{has the largest injection turbulence power, {more particles accelerate and} reach higher energies
than in  other models of Table~\ref{tabla}.} The differences increase with
energy, in consistency with Figure~\ref{fig:t-pinj}. Model III is the less
efficient as expected. Models I and II reach relativistic energies and exhibit a
high-energy power-law tail at  $t = 10^{0}$~$t_{\rm
A}$, while  model III only shows this behaviour at a later time.

\subsubsection{Dependence of the particles spectrum on $k_{\rm inj}$}

{We also compared the particles energy distribution of Models IV and V with Model I, differing only in the turbulence injection wavenumber $k_{\rm inj}$  (according to Table~\ref{tabla}),  for  the cases $c / V_{\rm A} = 50,\, 100$ and 150. As in Model I, the other models also exhibit {similar anisotropy} with  more particles being accelerated in the perpendicular  than in the parallel direction to the local magnetic field.  We also found  a weaker dependence in the evolved spectrum  with $k_{\rm inj}$  than with $P_{\rm inj}$, as before (see Figures~\ref{fig:t-pinj} and \ref{fig:t-kinj}). The differences between the distributions become more relevant at the high energies near the  saturation {of the fast growth region, with the distribution in Model I slightly flatter than {in} Models IV and V, respectively,  but within the error bars the distributions are very similar}.


\subsubsection{Effects of the initial particle energy distribution}

{In all the models studied so far,  we considered the same  initial thermal particle distribution   with a {thermal} speed $v_{\rm th} = 4 c / 100$.}
This distribution  corresponds to an
initial energy $E \sim 5 \times 10^{-4}$~$m_{p}c^2$.
{We have also computed  the acceleration of particles for a model with the same turbulence properties of model I and  with $V_{\rm A} / c  = 50$, but  changing the initial  particle distribution, considering two other cases:  one with a larger thermal speed ($v_{\rm th} = 4 c / 10$) and  the other with a smaller {one}   ($v_{\rm th} = 4 c / 1000$) than Model I. }
When comparing the acceleration rates {of these three cases we find  that after some time they all behave identically, i.e., the initially less energetic particle distribution takes longer time to enter the magnetic reconnection acceleration zone, but once the particles start to be accelerated, they evolve similarly to the other distributions, particularly } after reaching values of {the Lorentz factor $\gamma > 1$}.

Figure ~\ref{csnd} depicts the particles distribution after a  time, $t = 0.5~t_{\rm A}$ for the three cases. The best fitted total (normal + power-law) distribution is also shown for each model. We see that they all  produce a similar power-law, independently of the initial  thermal energy {of particles}.

\begin{figure}
\includegraphics[width=0.7\linewidth,angle=270 ]{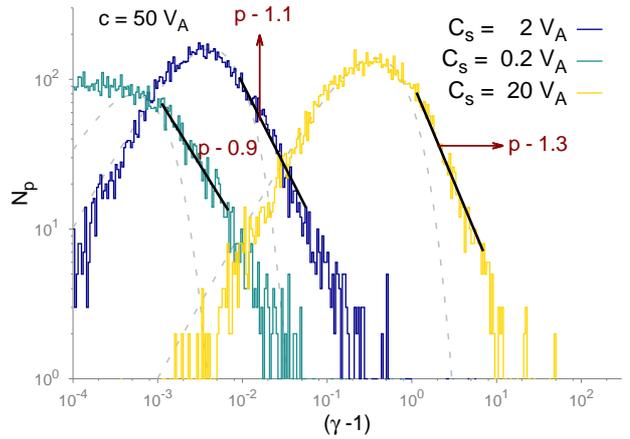}
\caption{Histograms of the number of particles for three different initial particle distributions for model I with $C_{\rm s} = 4 c / 10$, $4 c / 100$ and $4 c / 1000$. In all the cases $c / V_{\rm A} = 50$ and $t = 10^{-0.3}$~$t_{\rm A}$. }
\label{csnd}
\end{figure}

\section{Summary and Discussion}
\label{sec:concl}

In this work we have investigated the first-order Fermi acceleration of particles {within {large-scale} current sheets with fast magnetic reconnection driven by turbulence, using 3D collisional MHD simulations with the injection of test thermal particles,  following the same approach as in \citep{kowal12a}. We  extended  here this earlier study \citep[see also][]{kowal11,degouveia14,degouveia15} by examining the  effects of the parameters of the reconnection on  the effective acceleration rate and  the evolution of the spectrum of the particles.}
We considered models with different values of $V_{\rm A} / c $ and {different turbulence injection  scale and power.}

The main results can be summarized as follows.

\begin{itemize}

\item The acceleration time follows a power-law  dependence with the particle energy,  $t_{acc} \propto E^{\alpha}$, with 0.2 $<\alpha < 0.6$ {which is weakly sensitive to the magnetic reconnection parameters {of the injected turbulence}, tested for a large range of values  of  $ c / V_{A} \sim 20 - 1000$. }

\item The acceleration time {dependence with} the Alfv\'en velocity is $t_{acc}  \propto (V_{\rm A} / c)^{-\kappa}$,{ with $\kappa \sim 2.1 - 2.4$ {for particle kinetic} energies between  $E= (1 - 10^2)\,m_{p}c^2$, respectively and keeping the same trend approximately for larger energies (tested for model I).}

\item {For a given   value of} the $V_{\rm A}/c$ ratio, the acceleration time {is shorter for larger values of the  turbulence injection parameters,} i.e. $l_{\rm inj}$ and $P_{\rm inj}$, as expected from theory. Nonetheless,  the maximum differences between the models are generally less than  an order of magnitude and are within the error bars {due} to the uncertainties  in the evaluation of the acceleration times from the numerical simulations, {so that we can conclude that these dependences are not relevant.}

\item In all the cases studied here, the number of particles  being accelerated in the perpendicular direction to the local magnetic field  is larger than the ones being accelerated in the parallel direction. This unbalancing  is important to  ensure the effectiveness of the acceleration process (see below). 

\item The particle spectrum of the accelerated particles develops a high-energy tail, which can be fitted by a hard power-law index {$\propto$ $E^{p}$, with $p$ $\sim$ $-1.3$  to $-1$}
(or even a  little smaller)
 in the early times of the acceleration and is independent of the initial thermal energy of the injected particle distribution.

\end{itemize}


{These results}  have important implications for studies of particle acceleration specially in magnetically dominated regions of astrophysical environments like  the surrounds of GRBs,   black holes in AGNs and microquasars, and the relativistic jets associated to these sources.
{As remarked,} most studies of first-order Fermi particle acceleration by magnetic reconnection  have been performed  considering PIC  simulations   \citep[e.g.,][]{zenitani07,lyubarsky08, drake10,cerutti13,sironi14,guo14,guo15}, which apply only to the kinetic scales of the flow.
In order to probe the {large-scale} properties of the acceleration by magnetic reconnection beyond the kinetic scales in collisional astrophysical systems like those mentioned above,  an MHD description is required. The 2D and 3D studies undertaken by \citet{kowal11,kowal12a} and in this  work have explored exactly these macroscopic scales of the  acceleration by magnetic reconnection and thus are complimentary to the former kinetic studies.

It should be noticed also that, contrary to what has been found in PIC simulations \citep[e.g.,][]{guo14}, \citet{kowal11} demonstrated that the acceleration of energetic particles in 2D and 3D reconnection domains shows substantial differences, being more efficient in the second case. This justifies why our  analysis here  has focussed on 3D geometries of reconnection only.

{The earlier collisional numerical studies  (\citealt{kowal11, kowal12a}; see also \citealt{dmitruk03}) {and most of the ones presented here} have   neglected  the time evolution of the MHD environment. This is in general expected to be valid since this  time is much longer than the particle time scales, particularly when considering a first-order Fermi process in a statistically  steady state  turbulent domain. 
{In Sec.~\ref{time-evol}, we  explored  the particle acceleration considering different snapshots} of the reconnection domain  and found  no significant changes, as predicted.   Nonetheless, this  evolution may be important when considering more realistic non-steady flows and when calculating the spectra and loss effects  in real astrophysical systems. Preliminary steps in this direction have been performed in studies like, e.g. \cite{lehe09, khiali15a, khiali16, khiali15b}.}

{Earlier analytical studies of the first-order Fermi process in {large-scale} current sheets  \citep[e.g., GL05;][]{drury12} predicted that  the acceleration time would be  similar to that of shock acceleration,  and  the energy power-law spectrum of the accelerated particles could  be  
{either steeper or harder } than the one predicted for shock acceleration and 
nearly  independent on the reconnection velocity}.
These predictions, although based on rather simplified assumptions have been at least qualitatively  confirmed by the results of this work. 
For a broad range of reconnection velocities represented by a fiducial parametric space encompassing $V_{\rm A}/c \sim 1/1000- 1/20$, 
the acceleration time dependence with the {kinetic} particle energy is found to be $\propto E^{\alpha}$, with $\alpha \simeq 0.2-0.6$. 

Furthermore, the minimum analytically estimated acceleration time  according to Eq.~(\ref{eq:threshold}) is comparable to the values found in the simulations when the particles reach the maximum energy during} the  first-order Fermi acceleration in the reconnection  zone (the saturation energy). As we have seen, this maximum energy is attained when the particle Larmor radius becomes comparable to the size of the acceleration zone.

It is also remarkable {that}  the  power-law indices obtained for the particles distribution in the high-energy tail from our collisional MHD simulations in the large scales are  comparable to the values obtained from the PIC simulations in the kinetic scales of the plasma} \citep[e.g.,][]{zenitani01,drake13,sironi14,guo14,guo15,li15}.
\footnote{It should be stressed that, as in our model, the Fermi acceleration and resulting particle power-law spectrum obtained by \citet{guo14, guo15, li15} is due to the electric field produced by the magnetic fluctuations ($-{\bf u}\times{\bf B}$), while in the case of \citet{sironi14}, it is argued  that the acceleration is dominated by the resistive electric field component, which in our case is absent (see also \citealt{kowal12a}).}

We should stress that the acceleration process in magnetic reconnection sites with turbulence 
theory depends on $V_A$, $P_{\rm inj}$ and $l_{\rm inj}$)  that determines the first-order energy gain; (ii) the thickness of the turbulent region  which improves the particle scattering probability; and (iii) the strength and maximum scale of the velocity and magnetic field fluctuations within the turbulent region, which control the scattering mean free path (or time) which in turn depend on both $P_{\rm inj}$ and $l_{\rm inj}$. Therefore, the overall acceleration process is very complex. In this work we analysed  only the dependence of $V_{\rm rec}$ {with $V_A$ and the turbulence injection parameters. Both, $V_{\rm rec}$ and the acceleration efficiency are  clearly dominated by the $V_A$ dependence, as one should  expect for any  process driving the fast reconnection, though  we also obtained  some weak dependencies of the acceleration time with the turbulent parameters.} For instance, the scattering should happen at scales equal or smaller than $l_{\rm inj}$, 
this might be the reason why only the dependence on the injection power and not on injection scale is manifested at lower energies in our
results ($E_p < 10^2$, compare Figures~\ref{fig:t-pinj} and ~\ref{fig:t-kinj}). Moreover, at these scales particles can be scattered many times on the same
side of the current sheet, with the energy gain {temporarily} independent of the value of
$V_{\rm rec}$  until they are scattered back across the magnetic discontinuity again {to complete the first order Fermi cycle.}

{Having the points above in mind, we should remember that the turbulence is essentially the physical mechanism that drives fast reconnection in the {large-scale} current sheets studied here. This is a potentially  very important driving mechanism because turbulence is very common in astrophysical sources and environments. Nevertheless, the first order Fermi could in principle operate in current sheets with fast reconnection driven by other possible processes and the results should not {differ}  substantially {from} the present ones. This is compatible with the results found above that show only a weak dependence of the acceleration rate with the parameters of the turbulence. This may also explain  why our results are similar to those of kinetic PIC simulations, where the driving mechanisms of fast reconnection are generally very distinct.}

{It should be  also stressed that the collisional MHD  simulations shown here focussed on proton acceleration. Although  applicable to electrons too,  the numerical integration of the electron trajectories is much longer in MHD domains with test particles. Nevertheless, such tests are also needed. Hybrid simulations combining both the PIC and the MHD approach may be a good approach to this problem \citep[e.g.,][]{bai15}.}

We further remark that we have tried  to establish a link with the results of the PIC studies which probe only the kinetic scales up to 1000 skin depth scales. But  in our  collisional study  only the injected particles with Larmor radii near the MHD scales are effectively accelerated. This {limitation} can be also solved using hybrid codes able to resolve both the kinetic and the MHD scales {and make a smooth transition between them} \citep{degouveia15}.


{Another note is in order. This work should be distinguished from studies that  examined particle acceleration in pure turbulent environments (which are not embedded in {large-scale} current sheets, see e.g.,     
 \citealt{dmitruk03,zharkova11,dalena14,kowal12a,degouveia15,brunetti16}). For instance,  \citet{kowal12a}  have compared the two cases and concluded that in the cases with pure turbulence particle acceleration is probably dominated by a second order Fermi process, but further studies must be carried out in order to disentangle the processes.}

{Finally, cosmic-ray acceleration investigation in magnetic reconnection sites has still many challenges to overcome,  particularly in collisional MHD and relativistic regimes. The present study has tried to advance a little in the first of these topics. With regard to the second one, i.e., the study of acceleration in relativistic domains of reconnection, there has been some recent advances both in  collisionless descriptions \citep[e.g.,][and references therein]{cerutti13,sironi14,guo14,guo15}, and in collisional relativistic MHD fast reconnection involving  turbulence  (e.g., \citealt{degouveia15, lazarian15, singh16,takamoto15}; see also \citealt{degouveia14} for a short review of both approaches).
These are important issues to be explored further, specially for building more realistic models  of  flares and variability in the spectrum of compact sources to help in the interpretation of current  high energy  observations and in making  predictions for  upcoming new generation of instruments, like the Cherenkov Telescope Array \citep{cta11, cta13, sol13} and the ASTRI CTA Mini-Array \citep{astri15}.}


\section*{Acknowledgements}
M.~V.~d.V. acknowledges CNPq/Twas for financial support. E.M.G.D.P. acknowledges  partial support from the Brazilian agencies FAPESP (grant no. 2013/10559-5 and CNPq  (grant no. 300083/94-7). G.K. acknowledges support from FAPESP (grants no. 2013/04073-2 and 2013/18815-0) and PNPD/CAPES (grant no. 1475088).  M.~V.~d.V. thanks Reinaldo Santos-Lima for fruitful discussions on different topics addressed in this paper. The authors also acknowledge the anonymous referee for a careful review. This work has made use of the computing facilities of the Laboratory of Astroinformatics (IAG/USP, NAT/Unicsul), whose purchase was made possible by the Brazilian agency FAPESP (grant 2009/54006-4) and the INCT-A. M.~V.~d.V. thanks the great hospitality of the Instituto de Astronomia, Geof\'\i sica e Ci\^encias Atmosf\'ericas (S\~ao Paulo University), where part of this work was developed.









\bsp	
\label{lastpage}
\end{document}